\documentclass[eqsecnum, aps, pra, amsmath, amssymb, singlecolumn, superscriptaddress, 11pt ]{revtex4-2}

\usepackage{graphicx}
\usepackage{dcolumn}
\usepackage[colorlinks=true,
            linkcolor=blue,
            urlcolor=blue,
            citecolor=blue]{hyperref}
\usepackage{bm}
\usepackage{enumerate}
\usepackage{lipsum}
\usepackage{threeparttable}
\usepackage{epstopdf}
\usepackage{float}
\usepackage{xcolor,colortbl}
\usepackage[caption = false]{subfig}
\usepackage{tabularx}
\usepackage{chngpage}

\begin{document}


\title{Interference of longitudinal and transversal fragmentations in the Josephson tunneling dynamics of  Bose-Einstein condensates}
\author{Anal Bhowmik}
\email{anal.bhowmik-1@ou.edu}
\thanks{Present Address: The University of Oklahoma, Norman, Oklahoma 73019,USA}
\affiliation{Department of Physics, University of Haifa, Haifa 3498838, Israel}
\affiliation{Haifa Research Center for Theoretical Physics and Astrophysics, University of Haifa, Haifa 3498838, Israel}
\author{Ofir E. Alon}
\affiliation{Department of Physics, University of Haifa, Haifa 3498838, Israel}
 \affiliation{Haifa Research Center for Theoretical Physics and Astrophysics, University of Haifa, Haifa 3498838, Israel}

\date{\today}





\begin{abstract}
The dynamics of bosons in Josephson junctions have drawn much attention where the bosons are initially condensed. When interacting bosons tunnel back and forth along the junction,  depletion and eventually fragmentation develop. Here, we pose the question how do fragmented bosons  tunnel in a bosonic Josephson junction? To this end, we exploit the transverse degree-of-freedom of the junction to encode initial fragmentation to the bosonic cloud. We analyze the survival probability along the junction, fluctuations of particle positions across the junction, and the occupancy of the lowest single-particle states.  The dynamics found is rich and includes the speed up of the
collapse of density oscillations and slow down of the revival process. It is found that a fully fragmented state significantly accelerates the revival process compared to the conventional Bose-Einstein condensate. To explain the underlying many-body mechanism, we show that the initial fragmentation in the transverse direction interferes with the development of fragmentation in time along the junction. The dynamics of occupation in the first excited single-particle state defines whether interference of fragmentations occurs in the junction.  The interference mechanism is a purely many-body effect that does not occur in the mean-field dynamics.  All in all, we show that the interference of longitudinal and transversal fragmentations leads to   new rules for macroscopic tunneling phenomena of interacting bosons in traps.

\end{abstract}

\maketitle
\section{Introduction}
Atomic Bose-Einstein condensates (BECs) are a unique state of matter
and have been used as a flexible platform to explore a wealth of
physical phenomena  \cite{Dalfovo1999,Cederbaum2007, Alon2005, Bhowmik2016, Schmidt2022}.
A particular example of interest in the context of the present work is the Josephson effect \cite{Bloch2008}.  Over the years, various exotic features in  Josephson junctions of ultracold atoms, such as,   Josephson oscillations
\cite{Levy2007, Burchinati2017},  macroscopic self-trapping \cite{Smerzi1997, Albiez2005}, collapse and revival sequences \cite {Milburn1997}, 
 matter wave interferometry \cite{Schumm2005}, and squeezing \cite{Orzel2001, Wu2022}  have been investigated.   Josephson effects have been observed in complex systems, such as, spinor condensates \cite{Zibold2010}, polariton condensates \cite{Abbarchi2013}, fermionic superfluids \cite{Valtolina2015}, and spin-orbit coupled BECs \cite{Hou2018}.

In many-particle systems, interactions between the particles lead to correlations and their manifestation in a BEC is fragmentation which is a widely-studied  phenomenon \cite{Nozieres1982,Spekkens1999, Mueller2006, Bader2009, Zhou2013, Kang2014, Lode2017,Chatterjee2020}. The concept of fragmentation arises when   the reduced one-body density matrix starts to have more than one macroscopic eigenvalue  in contrast to the conventional BEC with just one macroscopically occupied state.  Hitherto, the development of fragmentation has been explored  in the atomic Josephson junction only for fully condensed states \cite{Sias2007, Sakmann2009,Erdmann2018,Theel2020,Vargas2021}. In recent years, the dynamics of Josephson junctions in two dimensions has been studied taking into account physics that emerges due to the transversal degree-of-freedom\cite{Fialko2012, Spagnolli2017,Burchianti2018,Xhani2020,Bhowmik2020, Bhowmik2022}.

 The transversal degree-of-freedom can also enrich the amount of initial  correlations in a BEC.
In the present context, fragmentation of  a bosonic system  in a two-dimensional setup can be generated due to the transversal degree-of-freedom alone.
This opens up the opportunity to ask and explore
how different degrees of fragmentation along the transverse direction would impact the Josephson-junction dynamics  in the longitudinal direction. 
More so, when initially condensed bosons tunnel back and forth (longitudinally) in a Josephson junction, they develop fragmentation.
Now, when we let (transversely) fragmented bosons  tunnel back and forth in the Josephson junction, would the initial fragmentation
impact the longitudinal fragmentation that develops in time in a Josephson junction?
So far, to the best of our knowledge, the 
possible interference of different fragmentations in a two-dimensional BEC has not been
discussed. 
The question is the following, can we combine the concepts of Josephson junction and different fragmentations  in a way   which would allow one  to have different internally correlated states?  This intriguing combination cannot be generated in one dimension and so far has not been explored.   It turns out that the macroscopic tunneling process becomes more  complicated when the different fragmentations interfere. 
Therefore, in this work, we propose a set-up to study tunneling processes involving both transversal and longitudinal fragmentations,
and the possible interference between them in time.
As we shall see below, the tunneling dynamics is significantly enriched.

With the emergence of fragmentation as a key perspective of quantum many-boson physics, we examine  the tunneling dynamics of a plethora of different  initially fragmented states, and observe an intricate  paradigm of tunneling dynamics taking into account the interference of fragmentations, to be defined precisely below.  We show that the impact   of transversal fragmentation on  longitudinal fragmentation   sets up  new rules of the  quantum tunneling problem. The rules of tunneling of a fragmented BEC manifests in quantum mechanical  quantities, namely, survival probability, build-up of   occupation  in the excited fragment,  and fluctuations of particles' positions.

\section{Theoretical setup}

The many-body Hamiltonian of $N$ interacting bosons in two spatial
dimensions reads: 
\begin{equation}
\hat{H}(\textbf{r}_1, \textbf{r}_2, . . . , \textbf{r}_N)=\sum_{j=1}^{N} \left[\hat{T}(\textbf{r}_j)+\hat{V}(\textbf{r}_j)\right]+\sum_{j<k} \hat{W}(\textbf{r}_j-\textbf{r}_k).
\end{equation}
 Here $\hat{T}(\textbf{r}_j)$ is the  kinetic energy of $j$-th boson and $\hat{V}(\textbf{r}_j)$ represents the trap  potential.   $\hat{W}(\textbf{r}_j-\textbf{r}_k)$  is the inter-boson interaction which is modelled as repulsive  Gaussian function \cite{Christensson2009,Doganov2013,Fischer2015} with  $W(\textbf{r}_j-\textbf{r}_k)=\lambda_0\dfrac{e^{-(\textbf{r}_j-\textbf{r}_k)^2/2\sigma^2}}{2\pi\sigma^2}$ where $\sigma=0.25\sqrt{\pi}$ is the width of the Gaussian function.   We take a Gaussian inter-boson interaction because in two spatial dimensions a delta-function potential does not scatter, see, e.g., \cite{Doganov2013}. The robustness of our findings to the range $\sigma$ is demonstrated in the   supplemental material \cite{Supplement}. Furthermore, we also demonstrate the robustness of the results to the shape of the inter-bosons interaction, taking dipolar interaction as a case study \cite{Supplement}.  Here $\lambda_0$ is the interaction strength which defines the interaction parameter $\Lambda_0=\lambda_0(N-1)$.   Throughout this work $\textbf{r}=(x,y)$  and the natural units $\hbar=m=1$ are employed. The number of bosons  considered  in this work is $N=10$.

\begin{figure}[!h]
{\includegraphics[scale=0.25]{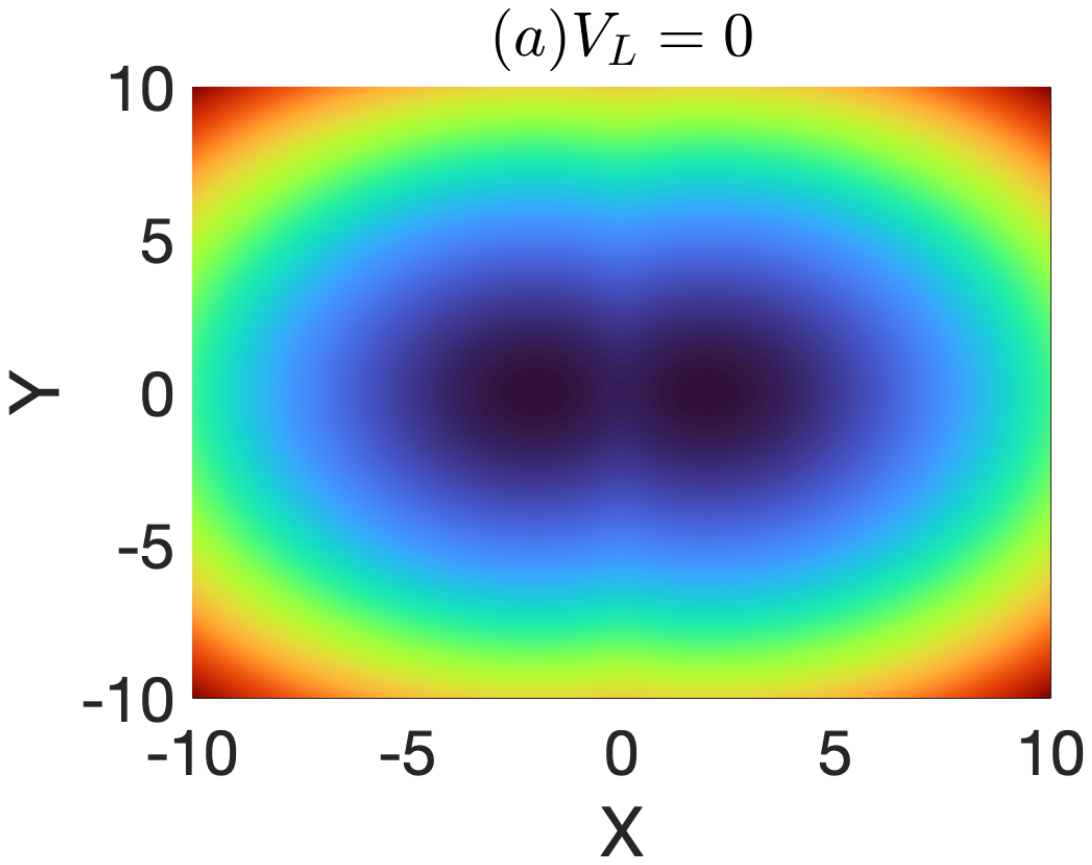}}
{\includegraphics[scale=0.25]{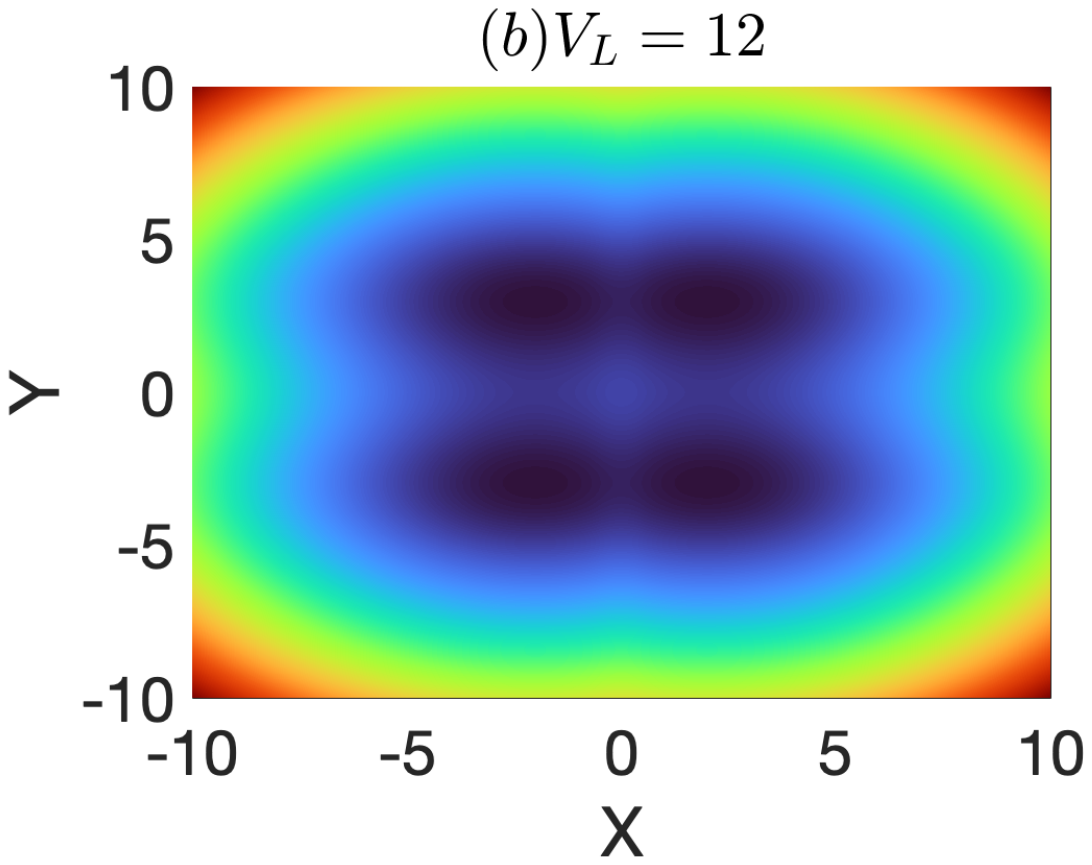}}
{\includegraphics[scale=0.25]{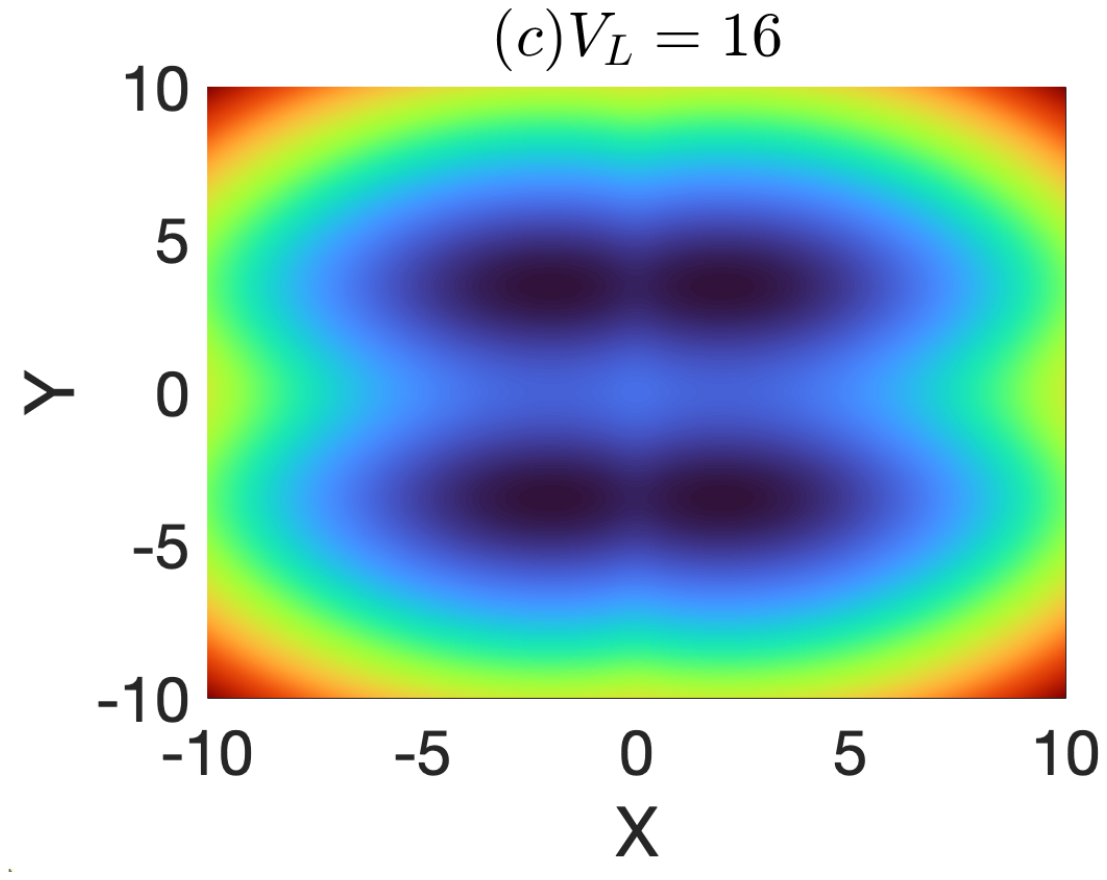}}\\
\vspace{-2cm}
\caption{ Trapping potential, $V^\prime(x,y)$, for the out-of-equilibrium dynamics.  The transversal barrier height  for the trapping potential is (a) $V_L=0$, (b) $V_L=12$, and  (c) $V_L=16$. Increasing  $V_L$ gradually, the trap,  where the dynamics takes place,  continuously transform from a double-well to  a four-well structure.  We show here dimensionless quantities.}
\label{Fig_new}
\end{figure}

The simplest trap one can imagine for our purpose is a double well of double wells. In order to create the different  fragmented ground states, we  design the one-body trap potential as  $V(x,y)=\dfrac{1}{2}(x+2)^2+V(y)$, where $V(y)=\dfrac{1}{2}y^2+V_Le^{-y^2/8}$, which transforms from single well to double well potential along the transverse direction  as $V_L$ is increased.  The  form of $V(x,y)$ manifests that the atoms are  initially trapped at the left side of space.  For the out-of-equilibrium tunneling dynamics, we quench the trap potential from $V(x,y)$ to   $V^\prime(x,y)$. The  form of $V^\prime(x,y)$ for the whole range of $y$ is  $\dfrac{1}{2}(x+2)^2+V(y)$ for $x<-\dfrac{1}{2}$, $\dfrac{1}{2}(x-2)^2+V(y)$ for $x>+\dfrac{1}{2}$, and  $\dfrac{3}{2}(1-x^2)+V(y)$ for $|x|\geq \dfrac{1}{2}$.   Here, the ramping up of $V_L$ leads to a four-well trap,   see Fig.~\ref{Fig_new} for selective barrier heights,  $V_L=0$, 12, and 16.  The quench of the trap potential described above is analogous to  the  quench of a standard bosonic Josephson junction, see, e.g. \cite{Sakmann2009, Bhowmik2020}.  Note that the longitudinal and transversal directions correspond to $x$- and $y$-directions, respectively. 

We make use of the powerful numerical many-body method, the bosonic version of the  multiconfigurational time-dependent Hartree  method \cite{Streltsov2007, Alon2008, Sakmann2009,Lode2017,  Nguyen2019, Lode2020},  also see its multi-layer version for  mixtures \cite{Kronke2013, Cao2013,Schurer2017,Chen2018,Mistakidis2019}. The method  incorporates quantum correlations exhaustively  to obtain an  in-principle numerically exact  ground states of different  initial fragmentations, and their  out-of-equilibrium dynamics.  To obtain the wavefunction,  a variationally optimal ansatz which is a linear combination of all permanents generated by distributing the $N$ bosons over $M$
time-adaptive orbitals is used. The  many-body wavefunction is described as \cite{Alon2008}
\begin{equation}\label{2.1}
|\Psi(t)\rangle =\sum_{\{\textbf{n}\}}C_\textbf{n}(t)|\textbf{n};t\rangle,
\end{equation}
where $C_n(t)$ are the expansion coefficients and $|\textbf{n};t\rangle= |n_1, n_2, . . ., n_M;t\rangle$. The number of time-dependent permanents $|\textbf{n};t\rangle$ is $\big(\begin{smallmatrix} N+M-1\\ N \end{smallmatrix}\big)$.   We recall that  for  $M= 1$  the many-body ansatz Eq.~\ref{2.1}  boils down to the time-dependent Gross-Pitaevskii equation \cite{Dalfovo1999}. As the number of  orbitals is increased, convergence of quantities, analyzed in this work, with $M$ is obtained. The theoretical method is well documented in the literature \cite{Lode2020}.
\vspace{-1.5cm}
In order to accurately capture the many-body physics,  we have performed the many-body computations with $M=8$ time-adaptive orbitals and the convergence is checked with $M=10$ time-adaptive orbitals (see the supplemental material \cite{Supplement}). For the numerical solution we use a grid of $128\times 128$ points in a box of size $[-10, 10)\times [-10, 10)$ with periodic boundary conditions. Convergence of the results with the number of grid points has been verified using a grid of $256\times 256$ points and  presented in \cite{Supplement}.

\section{Preparation and properties of the initial state}

\begin{figure}[!h]
{\includegraphics[trim = 0.0cm 1.0cm 0.1cm 1.5cm, scale=.25]{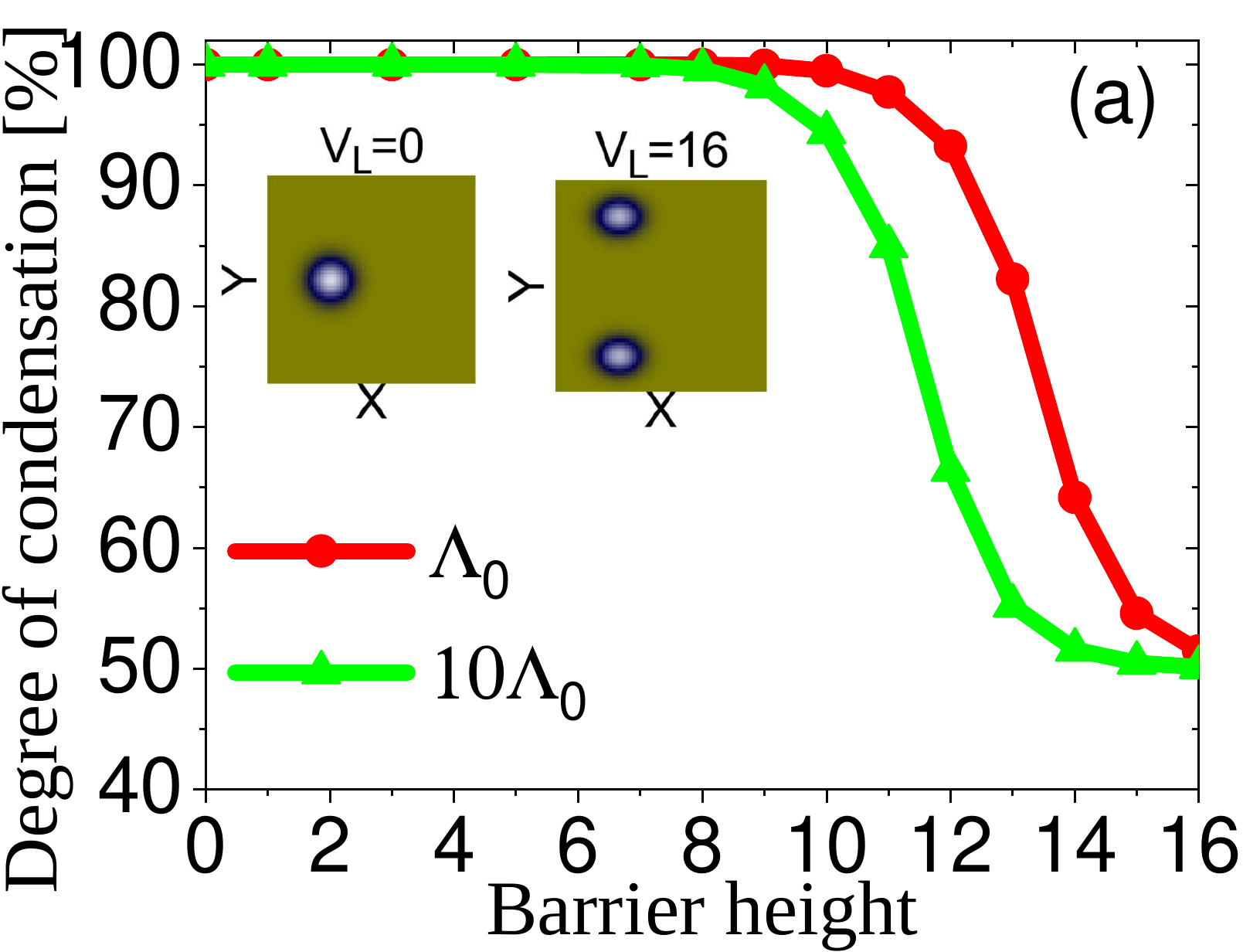}}
{\includegraphics[trim = 0.0cm 1.0cm 2.1cm 1.5cm, scale=.25]{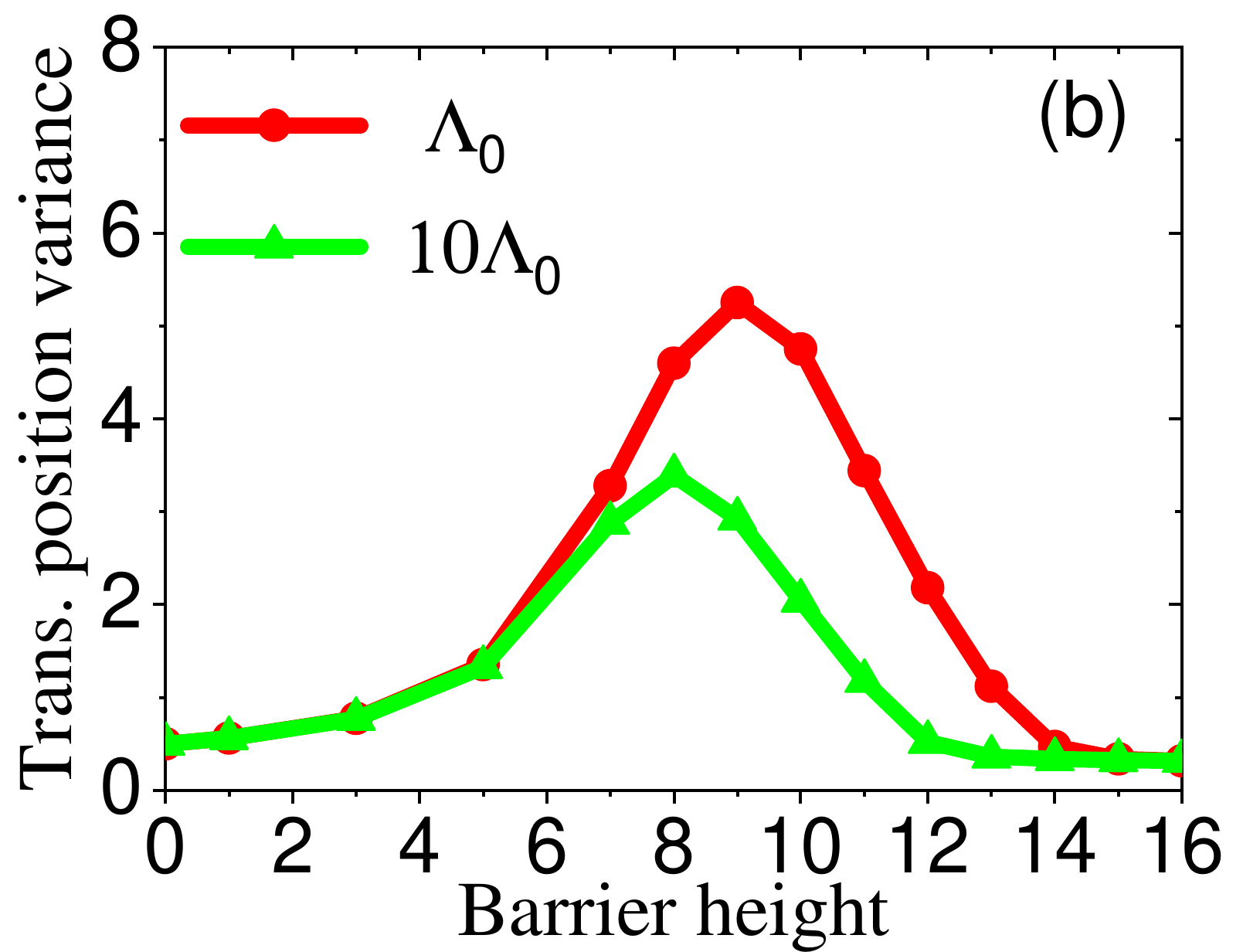}}\\
\caption{ Initial conditions. (a) Degree of condensation and (b) transversal position variance, $\dfrac{1}{N}\Delta^2_{\hat{Y}}$, as a function of the longitudinal barrier height ($V_L$) for two interaction strengths. The  inset in panel (a) shows the density per particle, $\frac{1}{N}\rho(\bf{r})$,  for the minimal and maximal $V_L$.  The number of bosons is $N=10$.   We show here dimensionless quantities.}
\label{Fig1}
\end{figure}

To study the tunneling dynamics of  fragmented BECs, we have to classify their basic properties depending on the degree of fragmentation.  In order to characterize  the condensed or fragmented state in the initial transversal double well, we present    the occupation of the first natural orbital, $n_1$, as a function of $V_L$. $n_1$   defines whether the ground state is condensed or fragmented in terms of the degree of condensation in the system. The  degree of condensation  is determined  from diagonalization of the reduced one-particle density matrix 
\begin{equation}
\rho(\textbf{r},\textbf{r}^{\prime})=N\int d\textbf{r}_2 . . . d\textbf{r}_N \Psi^*(\textbf{r}^\prime, \textbf{r}_2, ... ,\textbf{r}_N) \Psi(\textbf{r}, \textbf{r}_2, ... ,\textbf{r}_N),
\end{equation}
where $\rho(\textbf{r},\textbf{r}^\prime)=\rho(\textbf{r})$ is the density of the bosons. Fig.~\ref{Fig1}(a) depicts  the degree of condensation of the ground state  as a function of $V_L$ for two inter-boson interaction strengths $\Lambda_0=0.01\pi$  and  $10\Lambda_0$. The inter-boson interactions are weak in the sense that at  $V_L=0$ the ground state is more than $99.99\%$ condensed for $\Lambda_0$ and more than $99.9\%$ condensed for $10\Lambda_0$. As  expected, the  degree of condensation  of the ground state decreases with the barrier height, and  the ground state becomes $48.55\%$ fragmented for $\Lambda_0$ and $49.85\%$ fragmented for $10\Lambda_0$  when  $V_L=16$. We find that the ground state becomes two-fold fragmented from about $V_L\geq 7$, with the marginally occupied third and fourth orbitals having the  occupancy of  around $10^{-7}$,  and with the remaining four orbitals having  occupancy of  less than $10^{-7}$. The two-fold fragmented ground state implies  that two  natural orbitals are macroscopically occupied.  

For the considered geometry of the trap, the ground orbital is gerade ($g$-orbital) and the excited orbital is ungerade ($u$-orbital) along the $y$-direction.  In other words, the first fragment has $g$-symmetry and the second fragment has $u$-symmetry.  As fragmentation develops the $g$- and $u$-orbitals tend to be equally occupied. The topology of the investigation indicates that  parity in the $y$-direction is a good quantum number. Therefore,  we may call this fragmentation, developed by occupying the $u$-orbital along the $y$-axis,  the transversal fragmentation.

 Now, to analyze the  spatial distribution of bosons for different transversely fragmented ground states, we present  the many-body transversal position variance, 
\begin{equation}
\dfrac{1}{N}\Delta^2_{\hat{Y}}=\dfrac{1}{N}[\langle\Psi(t)|\hat{Y}^2|\Psi(t)\rangle-\langle\Psi(t)|\hat{Y}|\Psi(t)\rangle^2],
\end{equation}
in Fig.~\ref{Fig1} (b). The motivation is to use this quantity to analyze the dynamics of the initially transversely fragmented condensate as it tunnels back and forth along the Josephson junction, see section V.  The many-particle position variance is known to be sensitive to correlations \cite{Bhowmik2022, Klaiman2015}.  Initially,  $\dfrac{1}{N}\Delta^2_{\hat{Y}}$ monotonously grows with  $V_L$ from the initial value $0.5$, i.e., the value  for the harmonic potential at $V_L=0$,  and reaches  its maximal value at about $V_L=9$ for $\Lambda_0$ and  $V_L=8$ for $10\Lambda_0$.  Interestingly, due to the appreciable amount of the transversal fragmentation developed,  $\dfrac{1}{N}\Delta^2_{\hat{Y}}$ decays and tends to saturate when  $V_L$ reaches  the value 16. It is worthwhile  mentioning  that, unlike the many-body result, the mean-field $\dfrac{1}{N}\Delta^2_{\hat{Y}}$ monotonously grows
from the initial value $0.5$ with  increasing of the barrier height.  Also the 
 mean-field $\dfrac{1}{N}\Delta^2_{\hat{Y}}$ practically overlaps for $\Lambda_0$ and $10\Lambda_0$, see details in \cite{Supplement}.   It is noted that the many-body and mean-field variances in the longitudinal direction, $\dfrac{1}{N}\Delta^2_{\hat{X}}$,    are essentially  the same for all $V_L$  with value 0.5,  and are practically independent of the interaction strength chosen here  (and therefore need not be plotted). This suggests a very weak coupling between the $x$- and $y$-directions of the interacting bosons in the considered two-dimensional double-well trap for the initial conditions  for all $V_L$.

Summarizing, we have prepared initial states of different degrees of fragmentation in the transversal direction 
depending on the value of $V_L$. To investigate their out-of-equilibrium Josephson dynamics we propose to quench the trap to a four-well potential where the bosons can tunnel back and forth between the left and right transverse double wells. The main research question can be formulated now more precisely: Will  a transversely fragmented condensates develop longitudinal fragmentation in time? And if so, will the two fragmentations interfere with and impact each other?

 \section{Out-of-equilibrium  tunneling dynamics of a fragmented BEC: Emergence of interference}

\begin{figure}[!h]
{\includegraphics[scale=0.60]{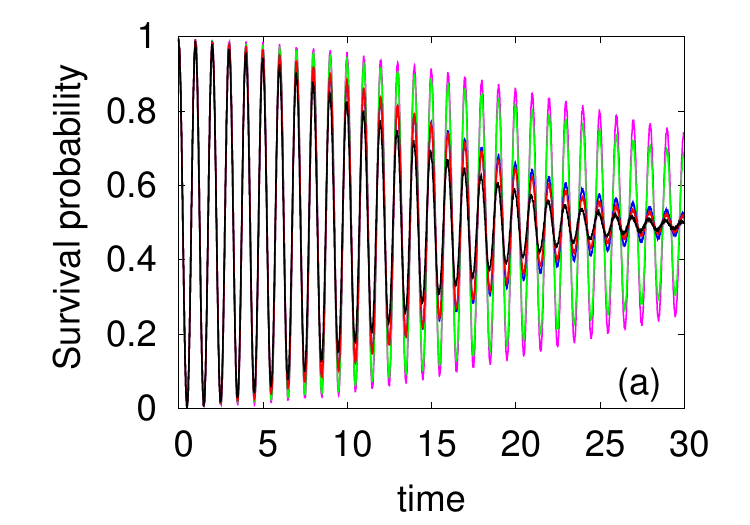}}
{\includegraphics[scale=0.60]{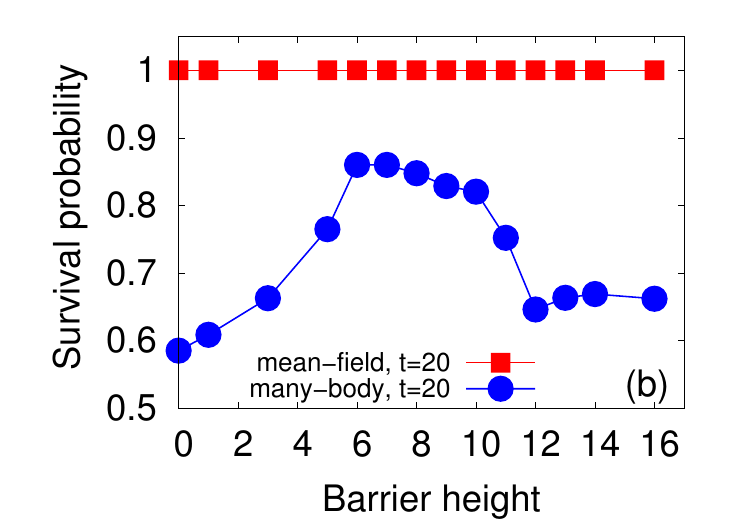}}\\
\caption{  Description  of tunneling dynamics using  many-body survival probability.   (a) Many-body survival probability in the left side of space, $P(t)$,  for the barrier heights  $V_L=0$ (black), 7 (magenta),  10 (green),  12 (red),  and 16 (blue) as a function of time.   For ease of  discussion, the least decaying to most decaying of $P(t)$ are for the barrier height $V_L=7$ (magenta), 10 (green), 16 (blue), 12 (red), and 0 (black), respectively. For the reader, we have separately plotted $P(t)$ for each barrier height in  Fig. S4 of the supplemental material. (b) $P(t)$ is plotted at $t=20t_{\text{Rabi}}$ as a function of the barrier height. The red squares and blue circles show the mean-field and many-body decay rate of $P(t)$, respectively. Unlike the  many-body dynamics, the mean-field results do  not distinguish the impact of the  transverse direction.  The  inter-boson interaction is  $\Lambda_0$ and the number of bosons $N=10$.  See the text for further details.   We show here dimensionless quantities.}
\label{Fig2}
\end{figure}

We begin investigating the tunneling dynamics with a  basic quantity, namely, the  survival probability in the left part of space, 
\begin{equation}
 P(t)=\int\limits_{x=-\infty}^{0}\int\limits_{y=-\infty}^{+\infty}d\textbf{r}\dfrac{\rho(\textbf{r};t)}{N},
\end{equation}
 where  $\rho(\textbf{r};t)$ is the time-dependent density.     In the many-body dynamics, one can find that  the frequencies of oscillations of $P(t)$  for different $V_L$ are essentially the same, see in Fig.~\ref{Fig2}, due to the practically same Rabi frequency along the $x$-direction \cite{Supplement}. Note that time is scaled by $t_{\text{Rabi}}$ in what follows where $t_{\text{Rabi}} (=132.498)$ is the time of a Rabi cycle \cite{Supplement}. The attractive feature lies in the decay rate of the amplitude of the many-body $P(t)$. We find that the amplitude of the many-body $P(t)$ decays in  time with  a different rate for different $V_L$. This decay occurs due to the development of longitudinal fragmentation in the tunneling process.   To distinctly identify the rate of decay  of $P(t)$ with different barrier heights, we present it    as a function of $V_L$ for a fixed time  in Fig.~\ref{Fig2} (b).  Here, we choose  a particular time $t=20 t_{\text{Rabi}}$  when the  bosons  tunnel back to the left side of space, and  is applicable for every even multiple of  $t_{\text{Rabi}}$. To compare $P(t)$ with the mean-field dynamics, we also plot the same at the mean-field level. Let us discuss first the mean-field dynamics of $P(t)$. As the trapping potential along the tunneling direction does practically  not change  with the barrier height, all the bosons continue to tunnel back and forth between the left and right parts of space at the mean-field level,  resulting in  "no-decay"  of  the density oscillations, see also \cite{Supplement} for the mean-field   $P(t)$. Therefore, in  the mean-field description of the dynamics, $P(t)$ is essentially invariant to the internal structure of the junction.

  Now, we come back to the details of many-body survival probability. The many-body dynamics of the system can be divided into three regimes  as a function of barrier height, i.e., lower barrier heights from $V_L=0$ to 6,   intermediate barrier heights from $V_L=7$ to 12, and higher barrier heights  from $V_L=13$ to 16.  Ramping up of the barrier from $V_L=0$ to 6, the initial ground state remains condensed  but  the decay rate  monotonously reduces 
 with $V_L$ as the  next transversal band gradually becomes closer to the lowest band.    Therefore in Fig.~\ref{Fig2} (b), 
 $P(t=20t_{\text{Rabi}})$  hits its maximal value for $V_L=6$. Once we pass $V_L=6$, the initial ground state is starting to lose its degree of condensation and it becomes a fragmented state, see Fig.~\ref{Fig1}(a). Therefore, more bosons are pushed to the $u$-orbital initially (at $t=0$) while we gradually increase the barrier height from $V_L=7$. The increased  initial occupancy in the $u$-orbital    leads to acceleration in the decay rate of $P(t)$ and thus the collapse of density oscillations speeds up. This gradual increase of the decay  rate  of $P(t)$ holds until the initial connection between the two orbitals ($g$-orbital and $u$-orbital) starts to be weaker for a certain internal structure of the junction. For the considered interaction parameter $\Lambda_0$, this particular  barrier height is $V_L=12$. The speeding up  of the collapse of the density oscillations  indicates that an intriguing and non-trivial effect occurs  between the longitudinal fragmentation developed during the tunneling process  and  the  initial transversal fragmentation.  This is a purely many-body  effect that  cannot be observed in the mean-field survival probability, see for details in \cite{Supplement}.  For $V_L>12$, the density collapse slows down again   until $V_L=16$, implying that the two orbitals of the ground state start to behave like two independent  states, and   the intriguing many-body effect  tends to diminish.
 
 By examining Fig.~\ref{Fig2} (b), one can find that the decay rate for  a fully condensed system, i.e., for $V_L=0$, is faster than for the fully fragmented system, i.e., for $V_L=16$. For the fully fragmented system, the initial occupancy is almost equally distributed between  the $g$-orbital and $u$-orbital. On top of that there is no connection between the two orbitals due to the combined effect of the inter-boson interaction and barrier height. Thus the two orbitals behave like two independent states tunneling back and forth, and each state consists of  almost $N/2$ bosons.  For a fixed interaction strength $\lambda_0$,  the decay rate of $P(t)$ is faster for a larger number of bosons, which explains the slower decay rate of a fully fragmented system compared to the fully condensed system. See the Appendix for the many-body  survival probability of $N=5$ and $N=10$ bosons in a two-dimensional double-well.

\begin{figure}[!h]
{\includegraphics[scale=0.60]{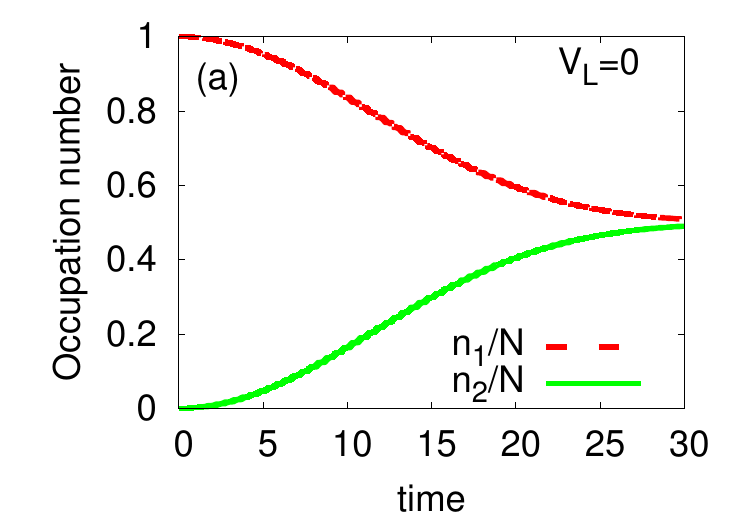}}
{\includegraphics[scale=0.60]{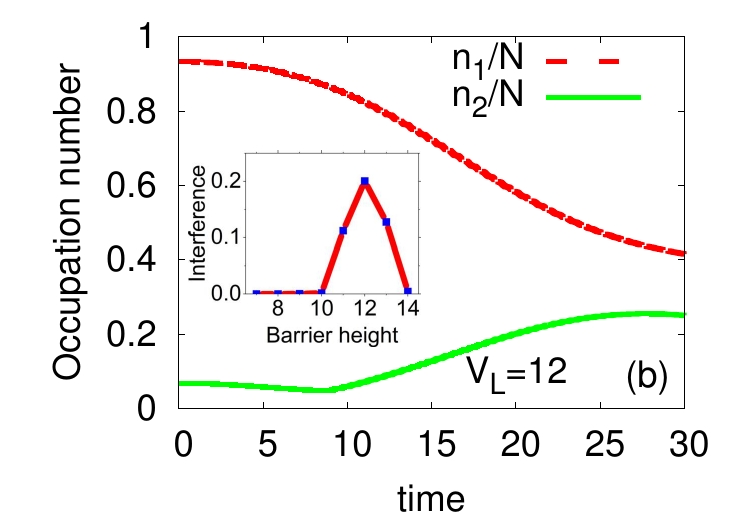}}\\
\vspace{0.5cm}
{\includegraphics[scale=0.60]{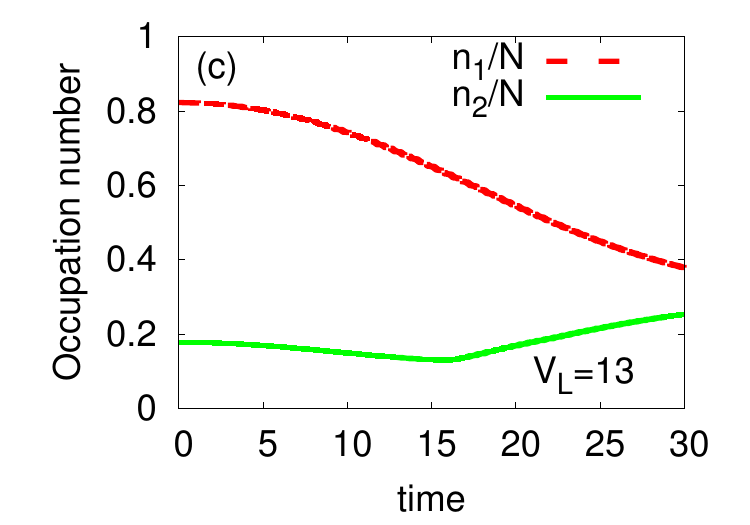}}
{\includegraphics[scale=0.60]{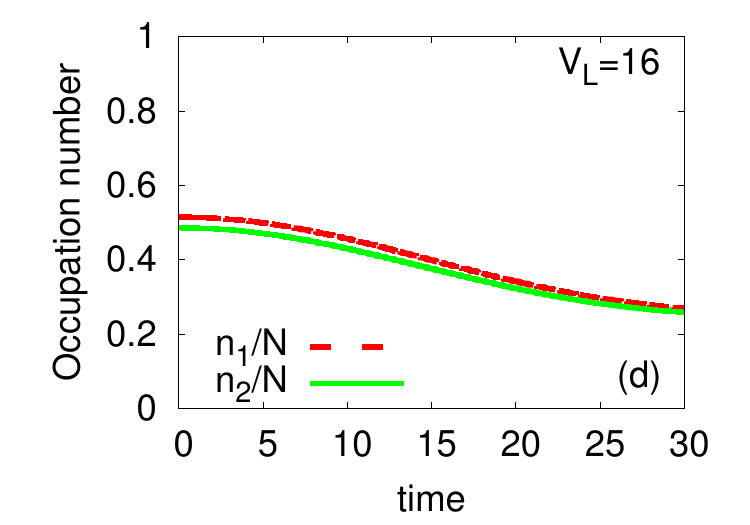}}\\
\caption{  Description  of tunneling dynamics using  the occupation numbers of the first two natural orbitals.    Occupations of the first and second natural orbitals, $\dfrac{n_1(t)}{N}$ and $\dfrac{n_2(t)}{N}$, respectively, for (a) $V_L=0$,  (b) $V_L=12$,  (c) $V_L=13$,  and (d) $V_L=16$.  The   traps for $V_L=0$,  $V_L=12$, and  $V_L=16$ are shown in Fig.~\ref{Fig_new}. The inset in panel (b) depicts the magnitude of  interference for the  barrier heights from $V_L=7$ to $V_L=14$.   The interference $[(n_u^{\text{maximal}}-n_u^{\text{minimal}})/N]$ occurs only for fragmented BEC  where $n_u$ is the occupation of $u$-orbital, provided $n_u^{\text{maximal}}$ is achieved at a later time than $n_u^{\text{minimal}}$.  The  inter-boson interaction is  $\Lambda_0$ and the number of bosons $N=10$.  See the text for further details.   We show here dimensionless quantities.}
\label{Fig2_new}
\end{figure}

\begin{table}[t]
\scriptsize
\caption {The symmetry of the first two maximally occupied orbitals   are highlighted for the barrier heights $V_L=0$ to 16. Here $t=0$ represents the initial condition and $t>0$ represents the out-of-equilibrium dynamics. For all barrier heights the first orbital remains as the $g$-orbital throughout the dynamical evolution (third column). On the other hand,  the symmetry of the   second orbital   is different depending on the barrier height and time of evolution (fourth column). For $V_L=0$ to $6$, the system is fully condensed and thus only one orbital, i.e., $g$-orbital, is  practically occupied at $t=0$, see also Fig.~\ref{Fig1}(a).  For $V_L\geq 7$, two orbitals are occupied at $t=0$ and they are $g$-orbital and $u$-orbital. Only for  $V_L=12$, 13, 15, and 16, the $u$-orbital remains as the second orbital throughout the dynamics. While for the  other barrier heights, the second orbital switches its nature between $u$-orbital and excited $g$-orbital or excited $u$-orbital.  As an example, for $V_L=7$,  the initial ($t=0$) second orbital is $u$-orbital,  then in the dynamics ($0<t\leq 0.20t_{\text{Rabi}}$) the second orbital remains as $u$-orbital,  and subsequently the second orbital becomes excited $g$-orbital for the time  window $0.20t_{\text{Rabi}}<t\leq 30t_{\text{Rabi}}$.  The number of bosons is $N=10$ and the interaction parameter $\Lambda_0$. For details see \cite{Supplement}.} 
\centering 
\begin{tabular}{c |c|c | c} 
  \hline
\hline
Barrier height 	& time &	 1st orbital	&	2nd orbital 		\\
\hline
$V_L=0$ to 6 & $t=0$& $g$-orbital & --\\
& $0<t\leq 30t_{\text{Rabi}}$& $g$-orbital & excited $g$-orbital\\
\hline
$V_L=7$  & $t=0$& $g$-orbital & $u$-orbital\\
& $0<t\leq 0.20t_{\text{Rabi}}$& $g$-orbital & $u$-orbital\\
& $0.20t_{\text{Rabi}}<t\leq 30t_{\text{Rabi}}$& $g$-orbital & excited $g$-orbital\\
\hline
$V_L=8$  & $t=0$& $g$-orbital & $u$-orbital\\
& $0<t\leq 0.81t_{\text{Rabi}}$& $g$-orbital & $u$-orbital\\
& $0.81t_{\text{Rabi}}<t\leq 30t_{\text{Rabi}}$& $g$-orbital & excited $g$-orbital\\
\hline
$V_L=9$  & $t=0$& $g$-orbital & $u$-orbital\\
& $0<t\leq 2.11t_{\text{Rabi}}$& $g$-orbital & $u$-orbital\\
& $2.11t_{\text{Rabi}}<t\leq 30t_{\text{Rabi}}$& $g$-orbital & excited $g$-orbital\\
\hline
$V_L=10$  & $t=0$& $g$-orbital & $u$-orbital\\
& $0<t\leq 0.76t_{\text{Rabi}}$& $g$-orbital & $u$-orbital\\
& $0.76t_{\text{Rabi}}<t\leq 30t_{\text{Rabi}}$& $g$-orbital & excited $g$-orbital\\
\hline
$V_L=11$  & $t=0$& $g$-orbital & $u$-orbital\\
& $0<t\leq 16.79t_{\text{Rabi}}$& $g$-orbital &  $u$-orbital\\
& $16.79t_{\text{Rabi}}<t\leq 16.96t_{\text{Rabi}}$& $g$-orbital & excited $u$-orbital\\
& $16.96t_{\text{Rabi}}<t\leq 17.20t_{\text{Rabi}}$& $g$-orbital &  $u$-orbital\\
& $17.20t_{\text{Rabi}}<t\leq 30t_{\text{Rabi}}$& $g$-orbital & excited $u$-orbital
\\
\hline
$V_L=12$ and 13 & $t=0$& $g$-orbital & $u$-orbital\\
& $0<t\leq 30t_{\text{Rabi}}$& $g$-orbital & $u$-orbital\\
\hline
$V_L=14$  & $t=0$& $g$-orbital & $u$-orbital\\
& $0<t\leq 25.70t_{\text{Rabi}}$& $g$-orbital & $u$-orbital\\
& $25.70t_{\text{Rabi}}<t\leq 30t_{\text{Rabi}}$& $g$-orbital & excited $g$-orbital\\
\hline
$V_L=15$ and 16 & $t=0$& $g$-orbital & $u$-orbital\\
& $0<t\leq 30t_{\text{Rabi}}$& $g$-orbital & $u$-orbital\\
\hline

\hline 
\end{tabular}
\label{table_1} 
\end{table}

Now,  to understand the underlying physics of the time-dependent  effect discussed above,  we  analyze further quantities, starting from   the  occupations of the natural orbitals.  Fig.~\ref{Fig2_new}  depicts the occupation of the first, $\dfrac{n_1(t)}{N}$, and second, $\dfrac{n_2(t)}{N}$, natural orbitals for $V_L=0$, $12$,  $13$,  and $16$.   To demonstrate how the occupation of the various orbitals affects the overall dynamics,  we have included the symmetries of the two maximally occupied  orbitals for all the barrier heights  from $V_L=0$ to 16  during the  dynamical evolution  in Table~\ref{table_1}.  It is found that $\dfrac{n_1(t)}{N}$ monotonously decreases, remains as $g$-orbital (see also section III for the initial orbitals), and eventually tends to saturate for all the initial states.  This holds true whether the system  is initially fully condensed, partially fragmented,  or fully fragmented,  see  Table~\ref{table_1}. Remarkably, the dynamical occupation of the second natural orbital shows an intriguing feature. We observe that when the  initial state is fully condensed at $V_L=0$ (or fully fragmented at $V_L=16$), $\dfrac{n_2(t)}{N}$ monotonously increases (or decreases). For the fully condensed system, the second natural orbital is an excited $g$-orbital and for the fragmented system  it is a  $u$-orbital,  see  Table~\ref{table_1}.    Interestingly, in Figs.~\ref{Fig2_new} (b) and (c), we find that, initially, the $u$-orbital loses its occupation  and thereafter  the  occupation  builds up in time. This transition, from   decreasing  of occupation  to building  up of  occupation  in the $u$-orbital represents the non-trivial effect which we call the interference of the longitudinal and transversal fragmentations, taking place for intermediate barrier heights from  $V_L=7$ to $V_L=14$.  Clearly, the interference of fragmentations occurs when all the following three conditions are satisfied chronologically, (i)   initially occupied  $u$-orbital, (ii) loss of  occupation  in the $u$-orbital, and (iii) build-up of  occupation  in the $u$-orbital.  Moreover, we observe that the $u$-orbital loses its 
  occupation until and unless there is a swapping of orders between higher natural orbitals, for details see \cite{Supplement}.

For completeness, we  have tabulated the  symmetry of the first two maximally occupied  orbitals  during the  time evolution, for all barrier heights  from  $V_L=0$ to 16, in Table~\ref{table_1}.        By inspecting  Table~\ref{table_1}, one can find that the  second maximally occupied orbital is changing its symmetry  during the time evolution  for the intermediate barrier heights  $V_L=7$, 8 , 9, 10, 11, and 14.  Among these intermediate heights, only for $V_L=11$  the nature of the second natural orbital  swaps between $u$-orbital and  excited  $u$-orbital. While for other intermediate barrier heights, the swapping takes place between   the $u$-orbital and excited  $g$-orbital, i.e., the symmetry changes. This swappings of orbitals are found by scrutinizing the occupancy of the orbitals with time, for details we refer to  \cite{Supplement}. Moreover,  for $V_L=11$, it is found that the nature of the second natural orbital is swapping multiple times between $u$-orbital and  excited  $u$-orbital. This swapping of orbitals showcases the intriguing dynamics of a fragmented BEC undergoing tunneling in the junction. 

 Now the magnitude of the interference between the longitudinal and transversal fragmentations is calculated from the difference between the maximal occupancy of the $u$-orbital after building up of  occupation  and its minimal occupancy when the transition from loss  to build up of  occupation  takes place, see the inset in Fig.~\ref{Fig2_new} (b).   Therefore, for fragmented BEC the  magnitude of interference is $\frac{n_u^{\text{maximal}}-n_u^{\text{minimal}}}{N}$, where $n_u$ is the occupation of $u$-orbital, provided $n_u^{\text{maximal}}$ is achieved at a later time than  $n_u^{\text{minimal}}$.  This analysis  identifies that the maximal interference of the longitudinal and transversal fragmentations takes place at $V_L=12$ for $\Lambda_0$.

 \section{Analysis of the interference of fragmentations}

\begin{figure}[!h]
 {\includegraphics[scale=.60]{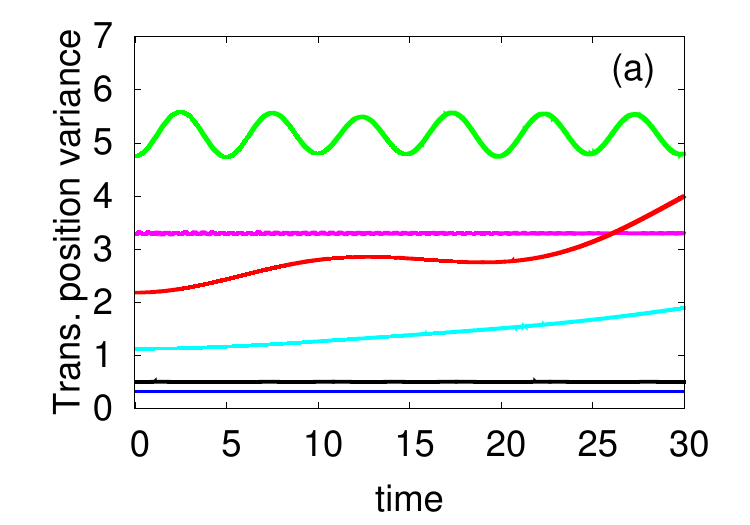}}
{\includegraphics[ scale=.60]{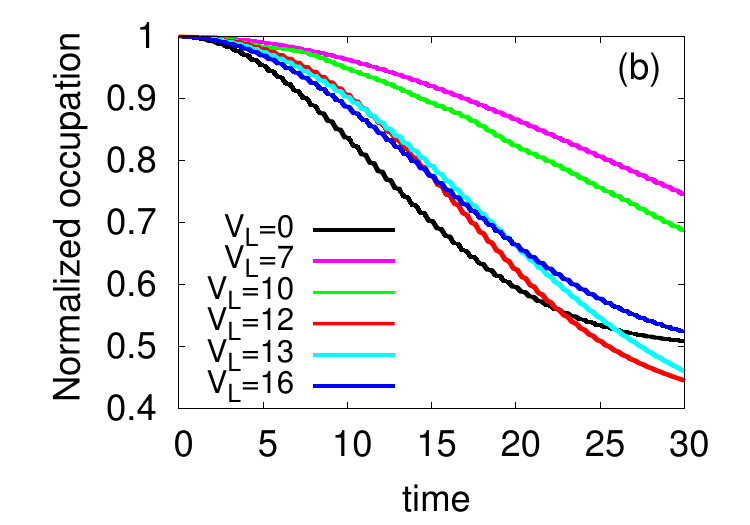}}
\caption{Interference of longitudinal and transversal fragmentations.  (a) Time-dependent transversal position variance per particle, $\dfrac{1}{N}\Delta^2_{\hat{Y}}(t)$, and (b) normalized occupation of the $g$-orbital, $\eta(t)={n_1(t)}/{n_1(0)}$.  The  inter-boson interaction is $\Lambda_0$ and the number of bosons $N=10$. Color codes are explained in  panel (b).  We show here dimensionless quantities. }
\label{Fig3}
\end{figure}

 Now, in order to show the consequences of  the interference of the longitudinal and transversal fragmentations on a quantum mechanical observable, we present the time evolution of the many-body $\dfrac{1}{N}\Delta^2_{\hat{Y}}(t)$ in Fig.~\ref{Fig3} (a).  As the inter-boson interaction is weak and since for $V_L=0$ the initial state is fully condensed, $\dfrac{1}{N}\Delta^2_{\hat{Y}}(t)$ is practically frozen   in time due to no interference between the longitudinal and transversal fragmentations.  At $V_L=7$, when the ground state is initially barely depleted (around 0.01\%, see section III),    $\dfrac{1}{N}\Delta^2_{\hat{Y}}(t)$ shows small oscillations during the tunneling process originating from  small interference of  fragmentations. Increasing of the initial fragmentation in the ground state, say, at $V_L=10$,  $\dfrac{1}{N}\Delta^2_{\hat{Y}}(t)$ shows oscillatory nature with constant amplitude and  frequency of oscillations, due to a stronger interference. Interestingly, for  $V_L=12$,  we find that $\dfrac{1}{N}\Delta^2_{\hat{Y}}(t)$ grows as time progresses. This growing nature of $\dfrac{1}{N}\Delta^2_{\hat{Y}}(t)$ signifies strong coupling of the $x$- and $y$-directions and it emerges from the   strong interference between the longitudinal and transversal fragmentations. For $V_L>12$, $\dfrac{1}{N}\Delta^2_{\hat{Y}}(t)$ slowly tends toward  essentially frozen dynamical behavior which exhibits decoupling of the  fragmentations.  Eventually, at $V_L=16$, we observe essentially constant dynamical behavior due to the practically null interference of the longitudinal and transversal fragmentations. Therefore, we conclude that the detailed investigation of the fluctuations in particle positions across the junction,  being a sensitive probe of correlations,  encodes the interference of   fragmentations.   All in all, we find that the oscillations of the  many-body $\dfrac{1}{N}\Delta^2_{\hat{Y}}(t)$   depend  significantly on the barrier height. Compared to the many-body dynamics, the mean-field $\dfrac{1}{N}\Delta^2_{\hat{Y}}(t)$ monotonously increases, showing  almost  frozen dynamics with fluctuation of  around $10^{-3}$,   for all barrier heights (see details in \cite{Supplement}).

Let us define a quantity to which we call the normalized occupation, $\eta(t)={n_1(t)}/{n_1(0)}$. We find that $\eta(t)$ would help us to compare the decrease in the first occupation number of systems with different degrees of initial fragmentation.   $\eta(t)$  also exhibits that the observed interference of fragmentations  varies in  time  during the  evolution. One can analyze the time-dependent nature of  the interference of fragmentations  by analyzing the real-time decay  of   the normalized occupation $\eta(t)$  of the $g$-orbital, see Fig.~\ref{Fig3} (b).   As one increases the barrier height from $V_L=0$ to $V_L=6$, the  rate of decay of  $\eta(t)$ decreases.   Further increase of the barrier height, until  $V_L=12$,  the rate of decay of  $\eta(t)$ increases   and, for $V_L>12$, it follows the previous trend of decreasing.  We examine the dynamics of $\eta(t)$ in time for different $V_L$. If we look into the details of  $\eta(t)$ for $V_L=12$, we notice that it crosses the corresponding $\eta(t)$'s found at $V_L=0,  13$, and 16 as time progresses, and that it exhibits the maximal rate of loss of   the degree of condensation. The behavior of  $\eta(t)$ suggests  that the interference between  the longitudinal and transversal fragmentations is indeed time dependent. By analyzing  Fig.~\ref{Fig3} (b), we find  the following rule of tunneling:  the interference between the  longitudinal and transversal fragmentations speeds up the rate of loss of   the degree of condensation $\eta(t)$  in the system.

\section{{Impact on  the revival process}}

So far,  we have dealt with the dynamics of various quantities  of a rather weakly interacting system  consists of  $N = 10$ bosons.   Now, to see the revival in reasonable time, we investigate the same problem  with stronger interaction, 10$\Lambda_0$.  As discussed in  Section III, for $V_L=0$,  the initial system ($t=0$) is about $99.99\%$  condensed for the  weaker interaction ($\Lambda_0$), and  for the  stronger interaction $(10\Lambda_0)$, it is about 99.9\% which is still rather condensed. These  values  give us a sense of how much stronger is  the interaction $10\Lambda_0$  compared to $\Lambda_0$. 

\begin{figure}[!h]
{\includegraphics[scale=0.60]{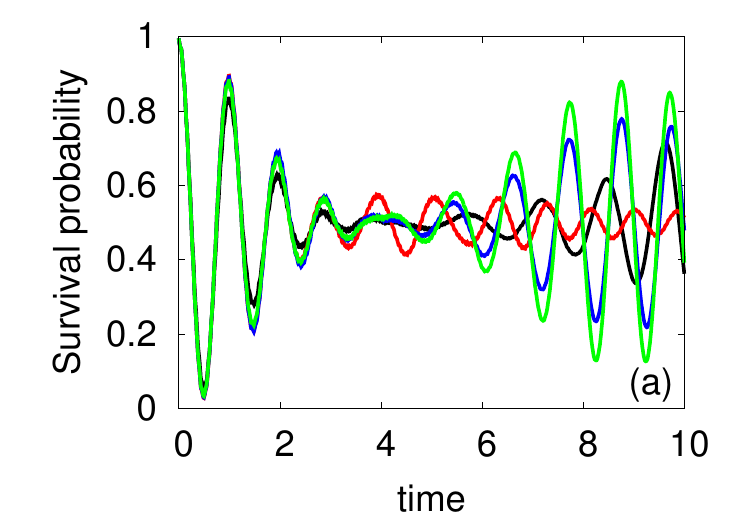}}
{\includegraphics[scale=0.60]{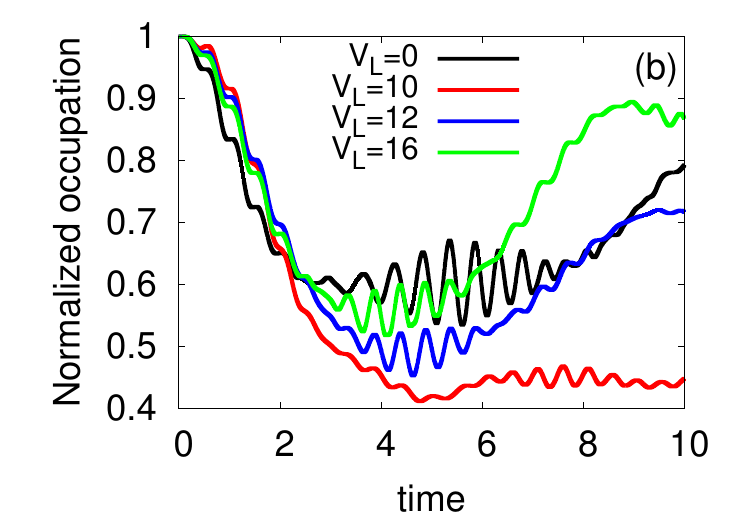}}\\
\caption{Tunneling dynamics for stronger interaction.  (a) Time-dependent survival probability in the left side of space, $P(t)$, and (b) normalized occupation  of the ground orbital, $\eta(t)={n_1(t)}/{n_1(0)}$.  The  inter-boson interaction is $10\Lambda_0$  and the number of bosons $N=10$. Color codes are explained in  panel (b).    We show here dimensionless quantities. }
\label{Fig4}
\end{figure}

 Here we would like to inspect what is the impact of interference of fragmentations, if any, on the revival process \cite{Milburn1997}. Fig.~\ref{Fig4} presents the time evolution of $P(t)$ and $\eta(t)$ for the  interaction parameter  $10\Lambda_0$.  Here,  $P(t)$ exhibits substantial richer  physics of the tunneling dynamics for  different  fragmented initial states. First of all, the density never tunnels  $100\%$  back  for all barrier heights, as the longitudinal fragmentation develops at the very beginning of the tunneling process due to the stronger interaction. In addition to the density collapse, also found for the weaker interaction $\Lambda_0$, here we observe a  different rate of revival of the density oscillations. We attribute this effect   
  to the interference of the initial transversal fragmentation  and the developed longitudinal fragmentation in the tunneling process.  We find that when the barrier height gradually increases  from $V_L=0$ to $V_L=10$,  the rate of revival of the density oscillations decreases, see the black and red curves of Fig.~\ref{Fig4}(a). Further increase of the barrier height, i.e., when $V_L>10$, the revival of the density oscillations increases and reaches  its maximal value,  when the  ground state is  initially fully fragmented.

The different rates of the revival process can be explained by the dynamics of $\eta(t)$, shown in Fig.~\ref{Fig4} (b). The general feature is that    all the ground states lose their  degree of condensation  with an oscillatory background  as time progresses. The maximal decay rate of $\eta(t)$, found at $V_L=10$, is the result of the strongest  interference of the longitudinal and transversal fragmentations, see \cite{Supplement}. Here we also find, as seen in the  case of weaker interaction,  that the rate of loss of  degree of condensation  of a fully fragmented BEC  is slower compared to the fully condensed state.  Moreover,  the rate of revival decreases when one moves from $V_L=0$ to $V_L=10$.  With further increase of $V_L$, $\eta(t)$ increases and becomes maximal for $V_L=16$.  

One finds out that    the many-body $P(t)$ and  the time-evolution of  $\eta(t)$ for stronger interaction exhibit a new rule for the revival dynamics of fragmented BECs:   the interference of the longitudinal and transversal fragmentations opposes the revival process.    Moreover, we observe that  a fully fragmented state, without  coupling between the longitudinal and transversal fragmentations, speeds up the process of revival compared to the conventional BEC.    As  discussed above, the fully fragmented BEC can be identified  as  two independent BECs with essentially $N/2$  bosons in each fragment.  Hence,   the observation  of revival dynamics of  fully condensed and fully fragmented systems  implies that, for a fixed inter-boson interaction strength $\lambda_0$, the revival process  takes place at a slower rate  for a BEC with a larger  number of bosons. To connect  further the revival dynamics 
 of a fully condensed system and a fully fragmented system, we present the survival probability  of $N=5$ and $N=10$ bosons  for the two dimensional double-well in the Appendix.

\section{Concluding remarks}

In conclusion, we explore the Josephson  dynamics of involved bosonic objects which undergo  a rich pathway from condensation to fragmentation  in a transversal double-well trap. We would like to bring out that such  intricate ground states shows new rules while tunneling.    We have demonstrated the physics behind the tunneling dynamics of the fragmented BEC by emphasizing  three limiting cases:  the first, when the ground state is initially fully condensed, the second, when the interference of the longitudinal and transversal fragmentations is maximal, and the third, when the ground state is initially fully fragmented with no coupling between the  fragmentations in the tunneling process. The interference of fragmentations in the tunneling process occurs when  the following three criteria are satisfied chronologically,  (a)  an initial occupation in the second fragment (the ungerade orbital along the $y$-direction),  (b)  the initial decrease of occupation   of the second fragment  in time, and (c)    the subsequent build up of occupation   in the second fragment in time.    The magnitude of the interference is defined by the difference between the maximal and minimal respective occupations of the second fragment.

 We have established that the interference of the  fragmentations constitutes  a new mechanism of the  tunneling process by analyzing the survival probability, details of fragmentation dynamics, and transversal position variance in the junction.  We find and explain how the interference of fragmentations governs the collapse and revival of the density oscillations and formulate general rules for macroscopic tunneling: (i) the interference between the longitudinal and transversal fragmentations speeds up the loss of  degree of condensation  in the junction, (ii) there is an optimal geometry  that maximizes the interference of  fragmentations while tunneling, (iii)  for a fixed inter-boson interaction strength $\lambda_0$, the loss of degree of condensation occurs at  a slower rate but the  revival process takes place at a faster rate for a smaller number of bosons in a BEC,  and   (iv) the interference of fragmentations delays  the revival process.

As the four-well set-up considered here is a minimal substructure of a two-dimensional optical lattice, tunneling of different quantum phases, such as, superfluid, Mott insulator \cite{Capello2007},  and fermionized Tonks–Girardeau gas \cite{Dunjko2001, Paredes2004}, can be studied by tuning the barrier height and the strength of inter-particle interaction.  Moreover, dipolar bosonic crystal orders and the dynamics of bosons in two-dimensional optical lattices, including the impact of competition between longitudinal and transversal fragmentations,  have the future scope to be investigated.   The many-body physics presented here could be relevant to the community working on atomtronics \cite{Amico2022} and metrology \cite{Baak2024}. Further investigations are warranted.

\section*{Acknowledgement}
This research was supported by the Israel Science Foundation (Grants No. 447/17 and 1516/19). AB acknowledges Sunayana Dutta for some helpful discussions.  AB also acknowledges Rhombik Roy for analyzing the results of revival dynamics and in preparing the Appendix.    Computation time on the High Performance Computing system Hive of the Faculty of Natural Sciences at University of Haifa  and the Hawk at the High Performance Computing
Center Stuttgart (HLRS) is gratefully acknowledged.

\section*{Appendix}

\begin{figure}[!h]
{\includegraphics[scale=0.60]{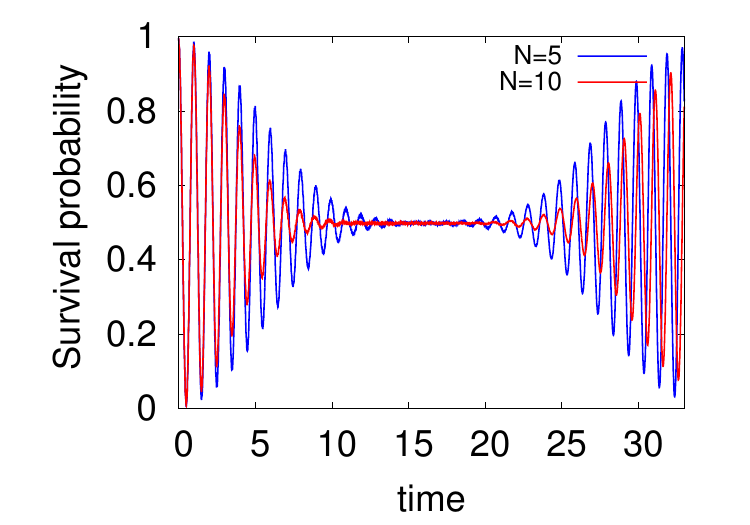}}
\caption{ Survival probability in  the left side of space, $P(t)$, for the barrier height $V_L=0$,  which describes a regular double-well potential in two dimensions. The inter-boson interaction is the same  as described in the main text.  The interaction strength is $\lambda_0=0.0111$ which corresponds to the interaction parameter $\Lambda_0=0.1$ for $N=10$ bosons. The results obtained  are for the numbers of bosons $N=5$ and $N=10$.  We show here dimensionless quantities. }
\label{Fig_App}
\end{figure}

 The appendix describes the tunneling dynamics in terms of the survival probability in  the left side of space, $P(t)$,     in a regular double-well potential in two spatial dimensions, i.e.,  for $V_L=0$.  The interaction strength is $\lambda_0=0.0111$ which corresponds to  the interaction parameter $\Lambda_0=0.1$ for $N=10$ bosons. The results are shown for $N=5$ and $N=10$ bosons, see Fig.~\ref{Fig_App}.  The results depict  the collapse of density oscillations and the revival dynamics. It is clearly seen that,  for a fixed inter-boson interaction strength   $\lambda_0$,  the decay rate of $P(t)$ is faster for $N=10$ while the revival time is faster for $N=5$. These results would help one  to demonstrate and discuss the  tunneling dynamics of a fully condensed system and a fully fragmented system presented in the main text. 

\clearpage
\section*{Supplemental material: Interference of longitudinal and transversal fragmentations in the Josephson tunneling dynamics of  Bose-Einstein condensates}

In this supplemental material, we promote the main text with further details. Here,  we start with  a brief mathematical description on the many-particle variance discussed in the main text. Next,  we present the initial-state  mean-field position variance as a function of barrier height.  Moving forward,  to represent and compare Josephson dynamics of different  fragmented states, we require a fixed time scale. As a useful information, we present the longitudinal and transversal Rabi frequencies in the four-well setup, considered in the main text, as a function of $V_L$.  Further, as a basic analysis tool of the Josephson dynamics presented here,  we provide the  mean-field results of the dynamics of the survival probability,  and  the longitudinal and transversal position variances. Moreover, in order to go deeper into the understanding of the interference in different fragmentation channels, we  demonstrate
the details of the fragmentation processes in terms of the occupancy of the higher natural orbitals and their  connection with  many-body longitudinal and transversal position variances. Next, we discuss the robustness of our results to the width of the inter-boson interaction potential. We also demonstrate the robustness of our results to  the shape of inter-boson interaction,  employing  the popular dipole interaction as a case study. Finally, we present the convergences of our results.

\clearpage

\section{Many-particle position variance}
The main text contains the results of the many-particle position variance which exhibits the impact of  interference of the longitudinal and transversal fragmentations developed during the tunneling process.   Here we present the mathematical formula of the many-particle position variance. The time-dependent variance per particle of an operator, $\hat{A}$,   is determined  by the combination of the expectation values of $\hat{A}$ and  $\hat{A}^2$. The expectation value of $\hat{A}$ $=\sum_{j=1}^N \hat{a}({r_j})$ depends only on  one-body operator. On the other hand,    ${\hat{A}}^2=\sum_{j=1}^N\hat{a}^2({r_j})+\sum_{j<k}2\hat{a}({r_j})\hat{a}({r_k})$  consists of one-body and two-body operators. Consequently,  the variance  is expressed as \cite{Alon2019b_s}
\renewcommand{\theequation}{S\arabic{equation}}
\setcounter{equation}{0}
\begin{eqnarray}\label{1}
\dfrac{1}{N}\Delta_{\hat{A}}^2&(t)&=\dfrac{1}{N}[\langle\Psi(t)|\hat{A}^2|\Psi(t)\rangle-\langle\Psi(t)|\hat{A}|\Psi(t)\rangle^2] \nonumber \\
&=& \dfrac{1}{N}\Bigg\{\sum_j n_j(t)\int d\textbf{r}\phi_j^*(\textbf{r}; t)\hat{a}^2({\textbf{r}})\phi_j(\textbf{r}; t)-\left[\sum_j n_j(t)\int d\textbf{r}\phi_j^*(\textbf{r}; t)\hat{a}({\textbf{r}})\phi_j(\textbf{r}; t)\right]^2 \nonumber \\ &+& \sum_{jpkq}\rho_{jpkq}(t)\left[\int d\textbf{r}\phi_j^*(\textbf{r}; t)\hat{a}({\textbf{r}})\phi_k(\textbf{r}; t)\right] \left[\int d\textbf{r}\phi_p^*(\textbf{r}; t)\hat{a}({\textbf{r}})\phi_q(\textbf{r}; t)\right]\Bigg\},
\end{eqnarray}
where $\{\phi_j(\textbf{r}; t)\}$ are the natural orbitals, $\{n_j(t)\}$  the natural occupations, and $\rho_{jpkq}(t)$ are the elements of the   reduced two-particle density matrix, 
\begin{equation}
\rho(\textbf{r}_1, \textbf{r}_2, \textbf{r}_1^\prime, \textbf{r}_2^\prime; t)= \sum \limits_{jpkq}\rho_{jpkq}(t)\phi_j^*(\textbf{r}_1^\prime; t) \phi_p^*(\textbf{r}_2^\prime; t)\phi_k(\textbf{r}_1; t) \phi_q(\textbf{r}_2; t).
\end{equation}
For one-body operators which are local in  position space, the variance described in Eq~\ref{1} reduces to \cite{Lode2020_s}
 
\begin{equation}\label{2}
\dfrac{1}{N}\Delta_{\hat{A}}^2(t)= \int d\textbf{r}\dfrac{\rho(\textbf{r};t)}{N}\hat{a}^2({\textbf{r}})-N\left[\int \dfrac{\rho(\textbf{r};t)}{N}\hat{a}({\textbf{r}}) \right]^2 + \int d\textbf{r}_1 d\textbf{r}_2 \dfrac{\rho^{(2)}(\textbf{r}_1, \textbf{r}_2, \textbf{r}_1, \textbf{r}_2; t)}{N}\hat{a}(\textbf{r}_1)\hat{a}(\textbf{r}_2).
\end{equation} 

\renewcommand{\thefigure}{S\arabic{figure}}

\setcounter{figure}{0}

\section{Mean-field  transversal position variance of the initial state}
\begin{figure*}[!h]
{\includegraphics[scale=.30]{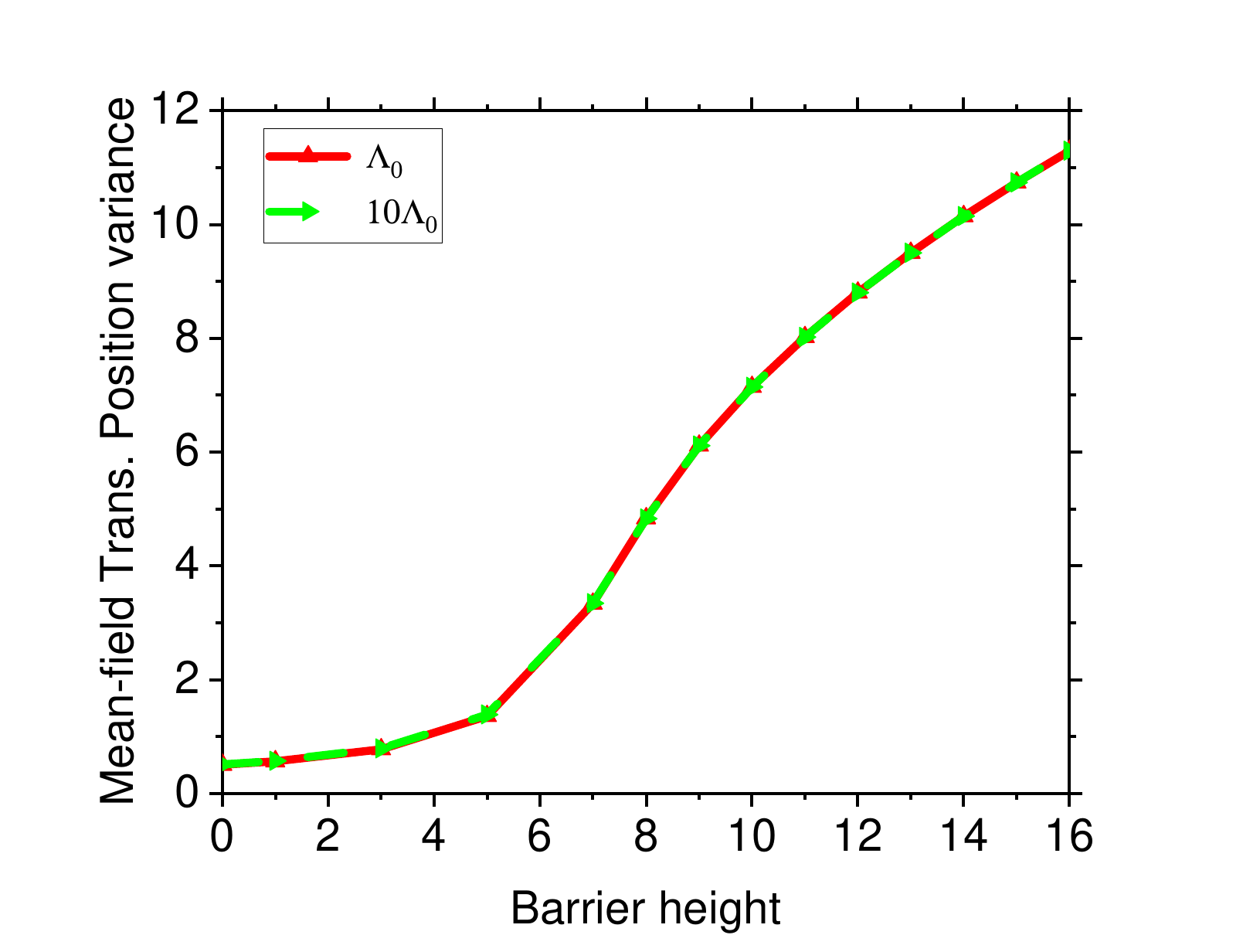}}\\
\caption{Mean-field transversal position variance, $\dfrac{1}{N}\Delta^2_{\hat{Y}}$, of the initial state as a function of the longitudinal barrier height ($V_L$) for two interaction strengths.  The number of bosons is $N=10$.   We show here dimensionless quantities.}
\label{FigS_new}
\end{figure*}

In the main text, we demonstrate the initial-state  many-body transversal position variance as a function of the barrier height. Here,  we investigate how the initial mean-field $\dfrac{1}{N}\Delta^2_{\hat{Y}}$ behaves with  $V_L$, see in Fig.~\ref{FigS_new}.  We observe that the mean-field $\dfrac{1}{N}\Delta^2_{\hat{Y}}$ monotonously   increases with the  barrier height which is unlike to the corresponding many-body result. Moreover, the mean-field $\dfrac{1}{N}\Delta^2_{\hat{Y}}$ is essentially independent of the interaction strength.

\section{Longitudinal and transversal Rabi periods for the four-well setup}
\begin{figure*}[!h]
{\includegraphics[scale=.30]{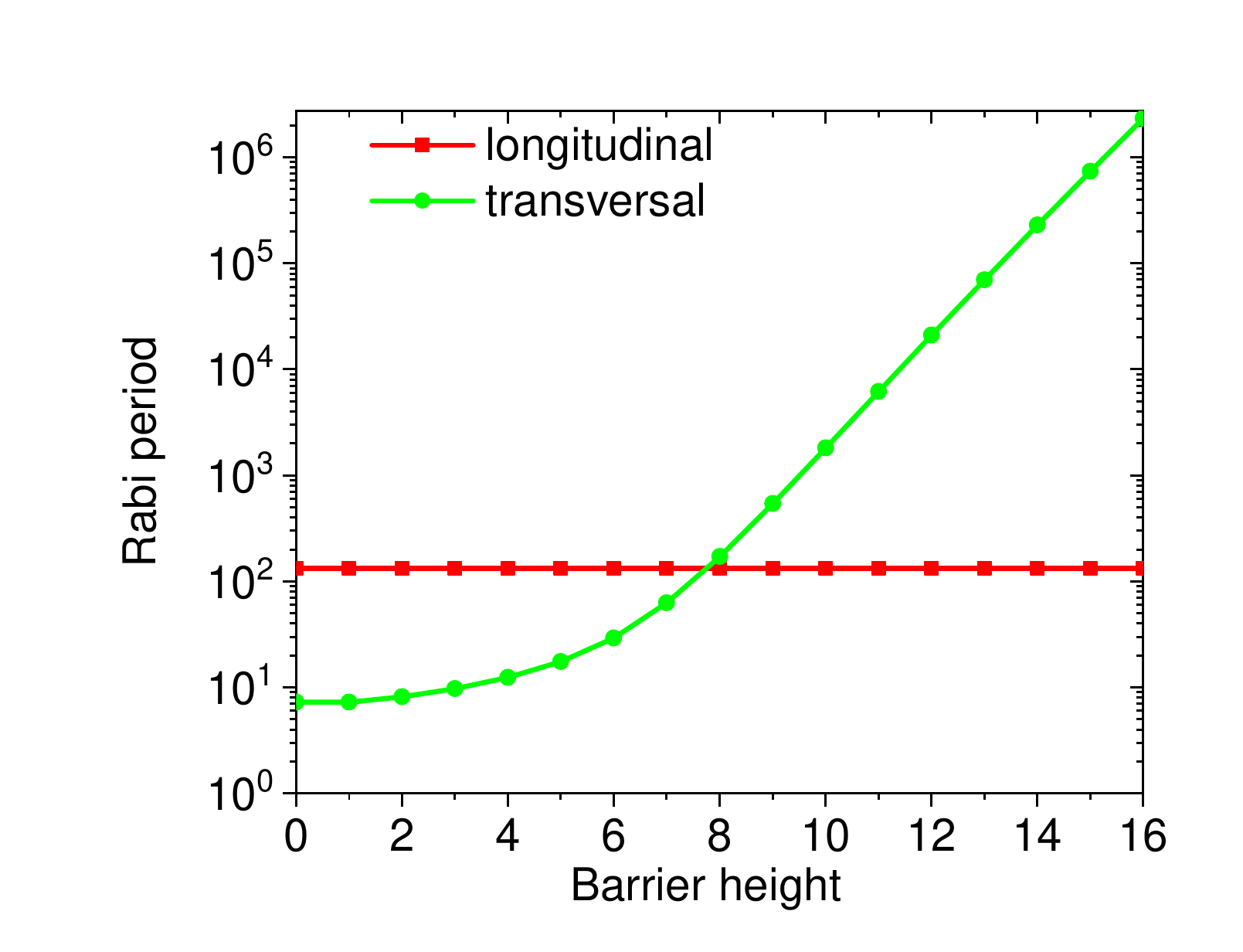}}\\
\caption{The Rabi periods (in logarithmic scale) along the longitudinal and transversal directions are computed  as a function of barrier height, $V_L$, for the four-well trapping potential in which the dynamics of the main text takes place.   We show here dimensionless quantities.}
\label{FigS1}
\end{figure*}

The tunneling of bosons, demonstrated in the main text, occurs for various barrier heights, $V_L$. Therefore,  to compare our results,   a natural choice of  time scale is required. Fig.~\ref{FigS1} provides the Rabi periods (in logarithmic scale) as a function of barrier height  along the longitudinal and transversal directions, $t_{Rabi}=\dfrac{2\pi}{E_X-E_0}$ and $t^{\prime}_{Rabi}=\dfrac{2\pi}{E_Y-E_0}$, respectively, where $E_0$ is the energy of the ground state, and $E_{X}$ and $E_Y$ represent the energies of the first excited states along the $x$- and $y$-directions, respectively. Here, the Rabi periods are calculated by diagonalizing the single-particle Hamiltonian using the discrete variable representation method. The  trapping potential  described  in the main text,  with the increasing barrier height,  changes from a double-well to a four-well potential.  As the one-body Hamiltonian is separable, it is noticed that $t_{Rabi}$ does not change with the barrier height with value 132.498 whereas $t^{\prime}_{Rabi}$ monotonously grows. $t^{\prime}_{Rabi}$ crosses $t_{Rabi}$ when $E_Y= E_X$  at $V_L\approx 8$. As we are interested in the tunneling dynamics along the longitudinal direction, we set the time-scale of the dynamics as $t/t_{Rabi}$. 

\section{No interferences between the  longitudinal and transversal   degrees-of-freedom within the mean-field dynamics}
\begin{figure*}[!h]
{\includegraphics[scale=0.6]{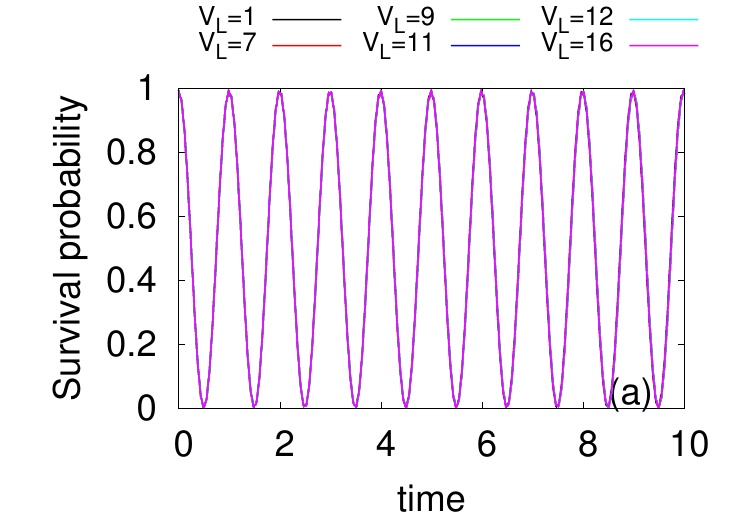}}
{\includegraphics[scale=0.6]{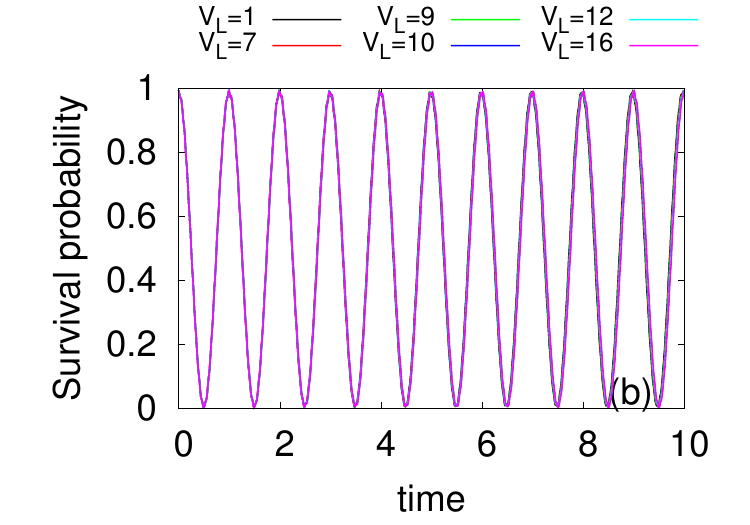}}\\
\caption{Time-dependent  mean-field survival probability in the left side of space, $P(t)$, for the interaction strengths (a) $\Lambda_0$ and (b) $10\Lambda_0$.  See the text for further discussion.  We show here dimensionless quantities.}
\label{FigS2}
\end{figure*}

We have seen in the main text that the amplitude of the  many-body survival probability, $P(t)$, decays for all barrier heights due to development of the longitudinal fragmentation. Here we show the dynamics of $P(t)$ as if we would have investigated  the tunneling phenomenon under mean-field theory. Fig.~\ref{FigS2} depicts the survival probability in the left side of space for the inter-particle interactions $\Lambda_0$ and $10\Lambda_0$.  Here, in  the mean-field dynamics, we observe that $100\%$ of the particles tunnel back and forth between the left and right parts of space with practically the same frequency of oscillations as a function of  the barrier height. Therefore, Fig.~\ref{FigS2} exhibits that the mean-field survival probability essentially does not  depend on  the shape of the initial density structure of the ground states, barrier height, and the considered inter-bosons interaction strengths. This is in sharp distinction from the many-body dynamics.  In other words, the mean-field dynamics show no interference  between the transversal and longitudinal   degrees-of-freedom. Hence, the simplicity of the mean-field dynamics may be used as a reference to define the interference of fragmentations at the many-body level of theory.

\section{Many-body survival probability}
\begin{figure*}[!h]
{\includegraphics[scale=0.6]{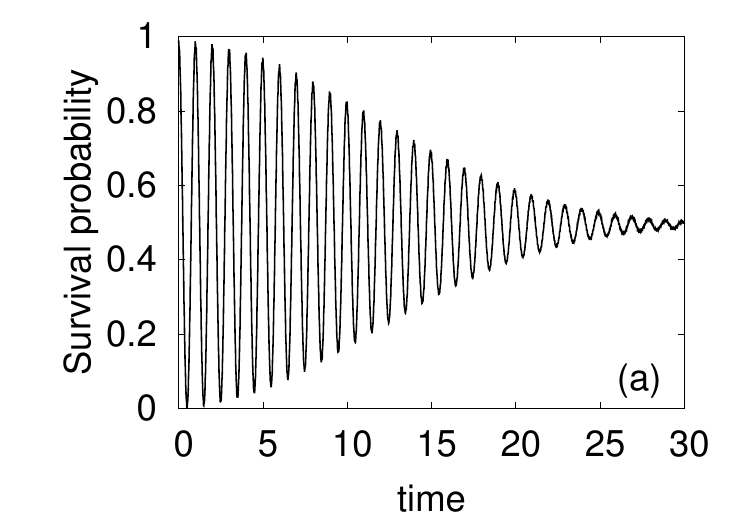}}
{\includegraphics[scale=0.6]{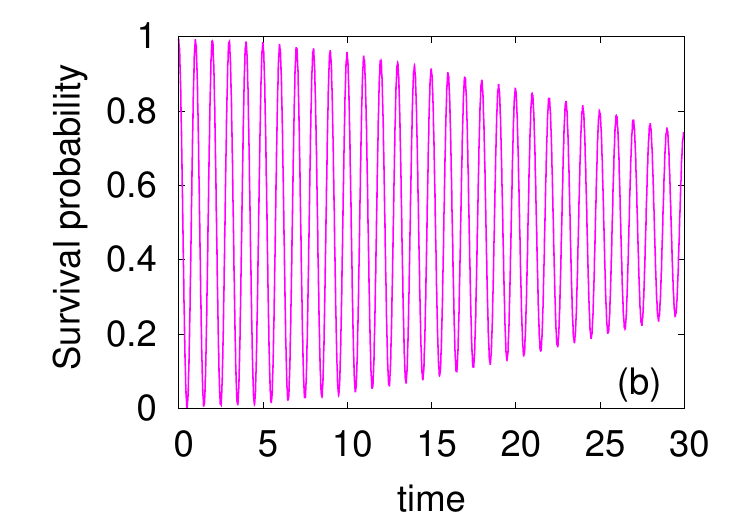}}\\
{\includegraphics[scale=0.6]{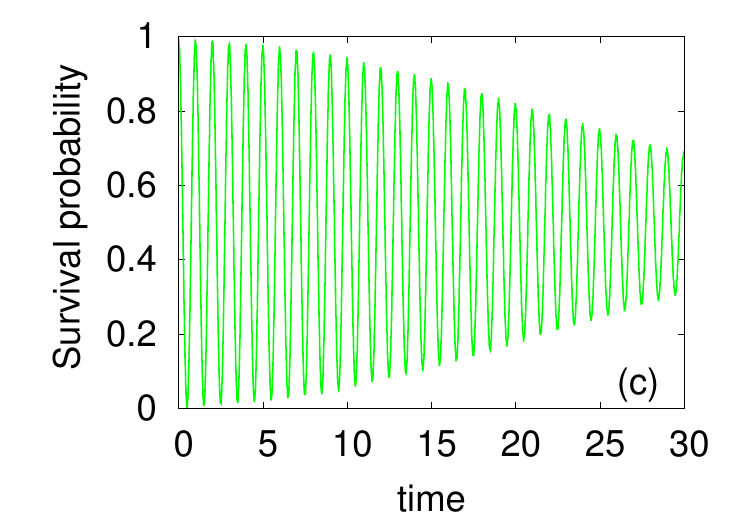}}
{\includegraphics[scale=0.6]{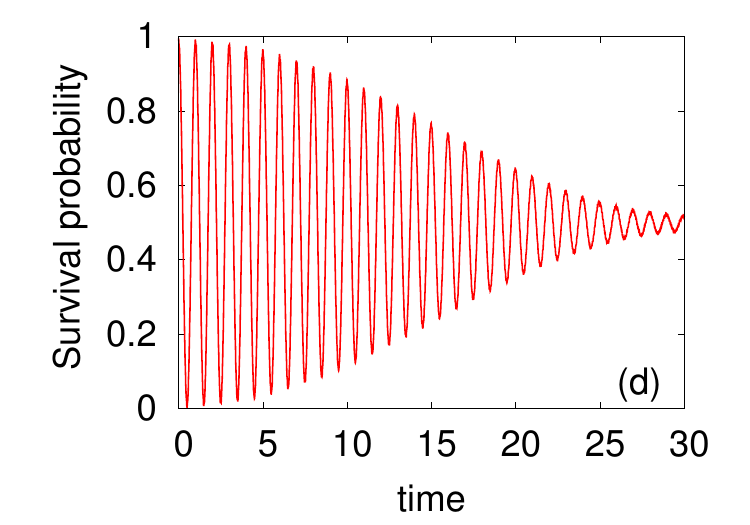}}\\
{\includegraphics[scale=0.6]{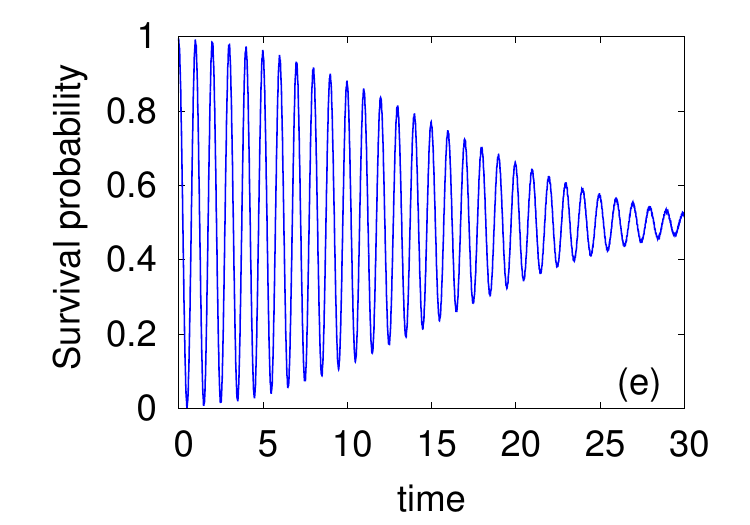}}\\
\caption{Time-dependent  many-body survival probability in the left side of space, $P(t)$, for the interaction strength  $\Lambda_0$ and number of bosons $N=10$. The barrier heights are (a) $V_L=0$, (b) $V_L=7$,  (c) $V_L=10$, (d) $V_L=12$, and (e) $V_L=16$. For comparison in a single plot, we refer Fig. 3(a) of the main text.  See the text for further discussion.  We show here dimensionless quantities.}
\label{Fig_surv_many}
\end{figure*}

The many-body survival probability in the left side of space are presented in  Fig.~\ref{Fig_surv_many} for the barrier heights $V_L=1$, 7, 10, 12, and 16. In the main text, we presented the  many-body survival probability for the mentioned barrier heights in a single plot to compare the decay rate of the survival probability and density collapse with respect to the   barrier height, see Fig. 3(a) of the main text.  Here, we separately plot the  many-body survival probability for each barrier height to get a better visibility for the reader.

\section{Details of the many-body fragmentations}
\begin{figure*}[!h]
{\includegraphics[scale=0.55]{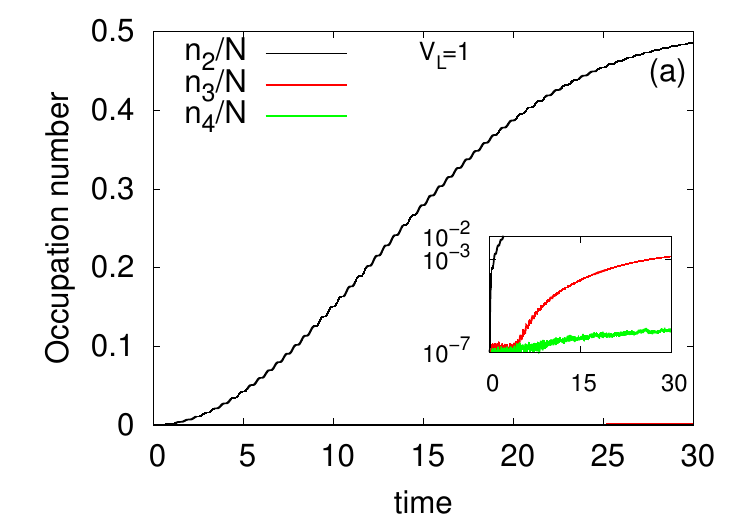}}
{\includegraphics[scale=0.55]{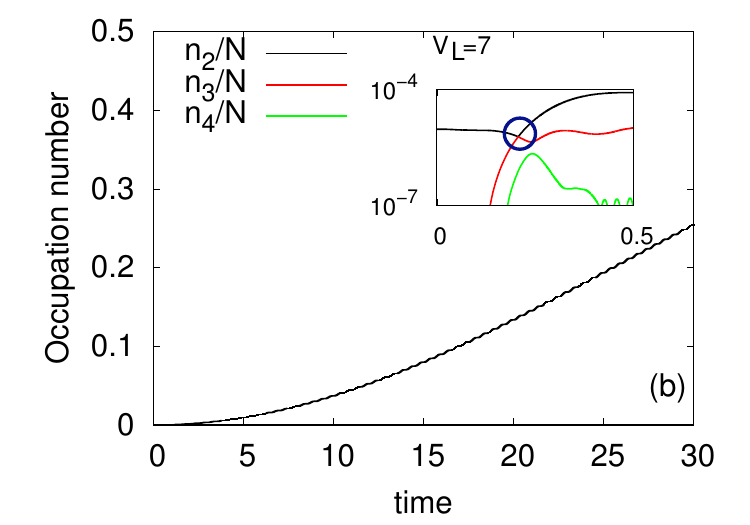}}\\
\vspace*{-0.3cm}
{\includegraphics[scale=0.55]{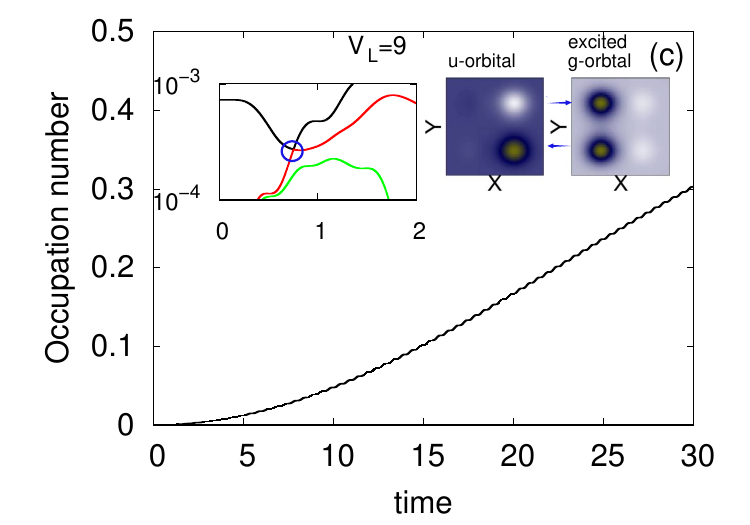}}
{\includegraphics[scale=0.55]{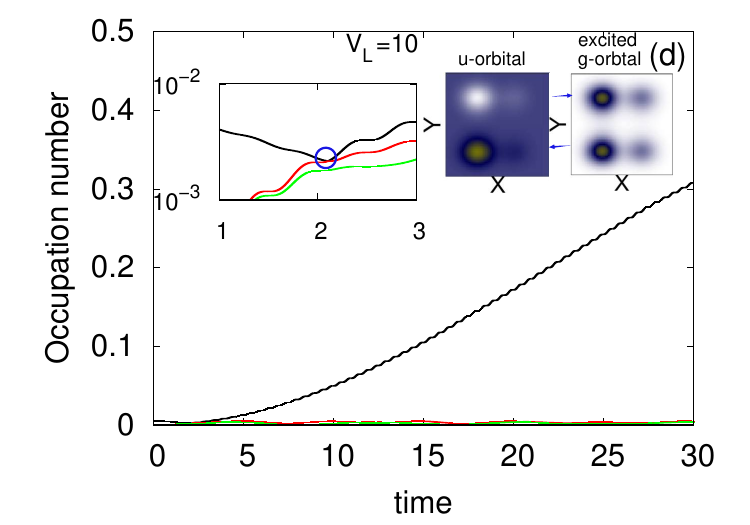}}\\
\vspace*{-0.3cm}
{\includegraphics[scale=0.55]{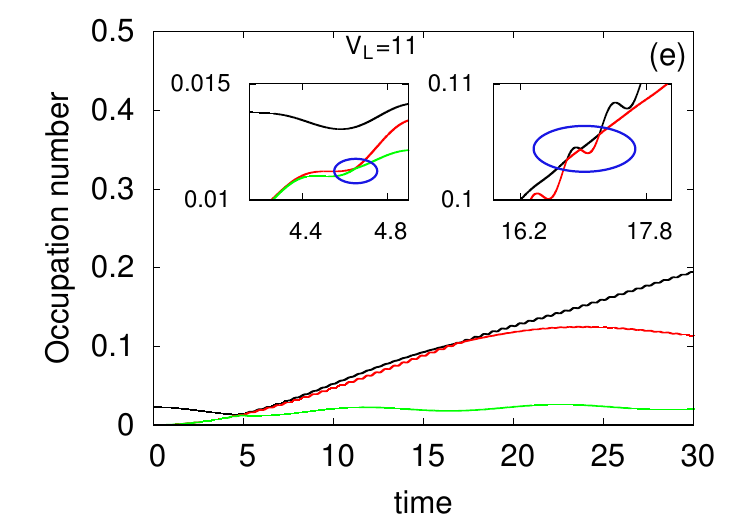}}
{\includegraphics[scale=0.55]{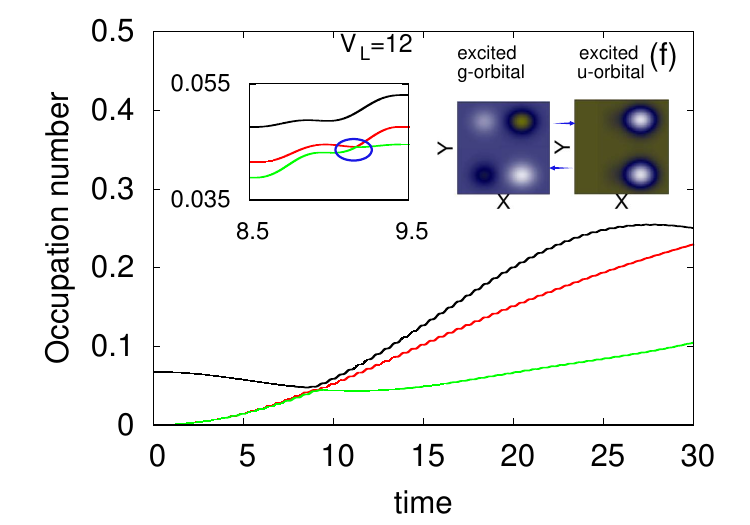}}\\
\vspace*{-0.3cm}
{\includegraphics[scale=0.55]{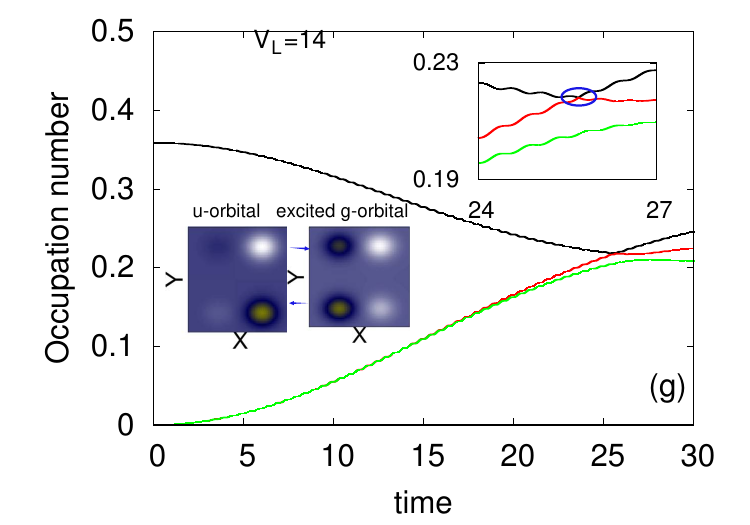}}
{\includegraphics[scale=0.55]{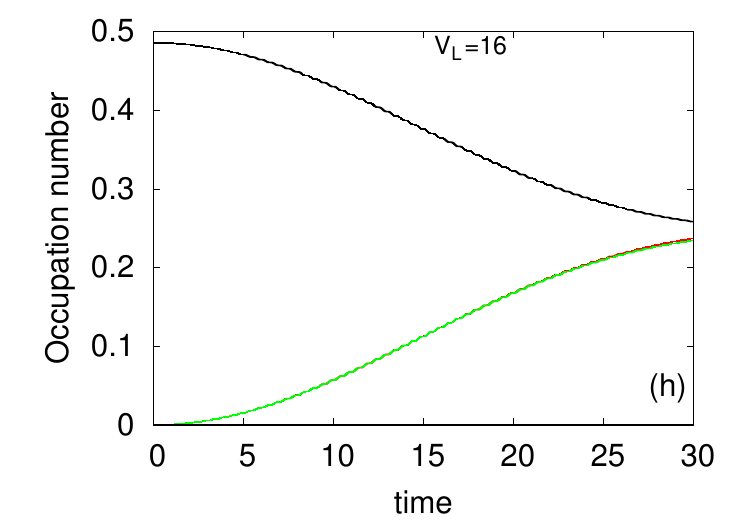}}\\
\caption{Details and mechanisms of the fragmentations (weak interaction). Time evolution of the occupation numbers per particle of the higher natural
orbitals, $\dfrac{n_{{j=2,3,4}}(t)}{N}$, for different barrier heights. The  inter-boson interaction is $\Lambda_0$   and the number of bosons $N=10$. The insets of panels (a) to (g) magnify the same plots. In the insets of panels (b) to  (g), we also present swapping of orders of the orbitals. This swapping of orbitals happens when the $u$-orbital shows a transition from loss of coherence to build up of coherence. To guide the eye, we mark the time of swapping of orders of the orbitals  with a blue circle.   Color codes are explained in panels (a) and (b).   We show here dimensionless quantities.}
\label{FigS3}
\end{figure*}

\begin{figure*}[!h]
{\includegraphics[scale=0.6]{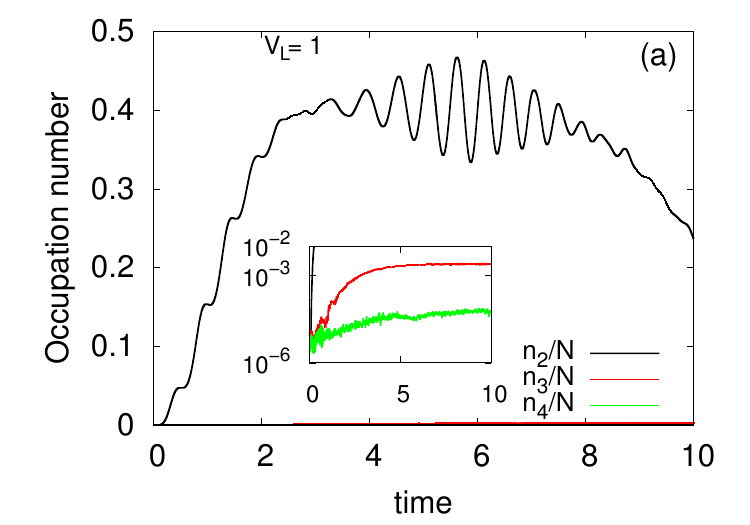}}
{\includegraphics[scale=0.6]{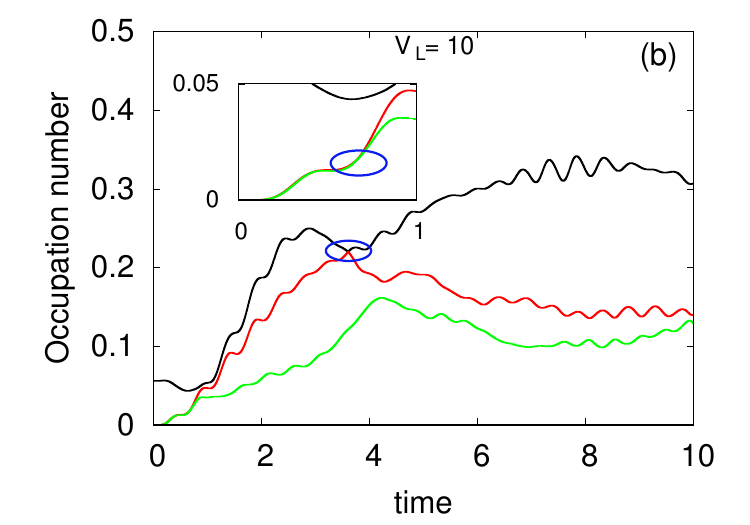}}\\
\vspace{0.2cm}
{\includegraphics[scale=0.6]{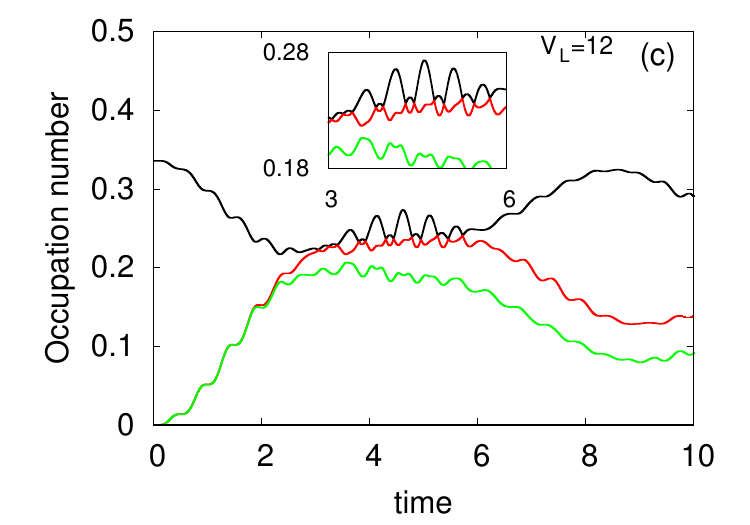}}
{\includegraphics[scale=0.6]{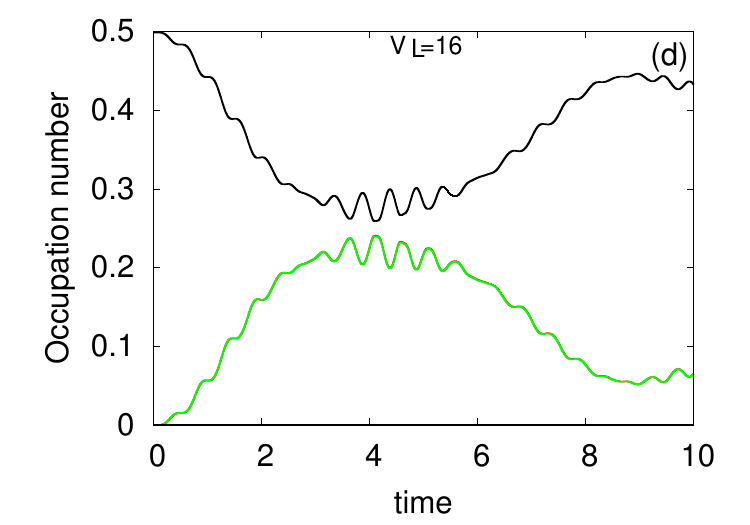}}\\
\caption{Details and mechanisms of the fragmentations (stronger interaction). Time evolution of the occupation numbers per particle of the higher natural
orbitals, $\dfrac{n_{{j=2,3,4}}(t)}{N}$, for different barrier heights. The  inter-boson interaction is $10\Lambda_0$   and the number of bosons $N=10$. The insets of panels (a) to (c) magnify the same plots. In the inset of   panel  (b), we  present swapping of orders of the orbitals. This swapping of orders happens when the  $u$-orbital shows a transition from loss of coherence to build up of coherence. To guide the eye, we mark the time of swapping of orbitals  with a blue circle.   Color codes are explained in panel (a).   We show here dimensionless quantities.}
\label{FigS4}
\end{figure*}

The main text shows that the occupation of the first natural orbital decreases with time,  which implies the  growing occupations of the higher natural orbitals.  Now, we examine and compare the microscopic mechanism of how the higher natural orbitals, $\dfrac{n_{{j=2, 3, 4}}(t)}{N}$, become populated as a function of $V_L$. Figs.~\ref{FigS3} and \ref{FigS4} depict the occupations of the most dominant higher natural orbitals, i.e., the second, third, and fourth natural orbitals for the interaction strengths $\Lambda_0$ and $10\Lambda_0$, respectively. 

Let us start with the discussion of Fig.~\ref{FigS3}. If one gradually moves from $V_L=1$ to $6$, the initial state continuously deforms yet  maintaining its coherency. The deformation of the initial ground state delays the process of losing the coherence in the dynamics.  Therefore, from $V_L=0$ to $V_L=6$, although  all the higher natural orbitals become occupied with time, the rate of occupancy of the second natural orbital is faster for  $V_L=0$ compared  to $V_L=6$. From $V_L=0$ to $6$, the second, third, and fourth natural orbitals are excited $g$-orbital,   $u$-orbital, and excited $u$-orbital, respectively, discussed in the main text. The order of populations  of the orbitals is preserved throughout the dynamics until $V_L=6$, i.e., excited $g$-orbital$>u$-orbital$>$excited $u$-orbital.

\begin{figure*}[!h]
{\includegraphics[scale=0.4]{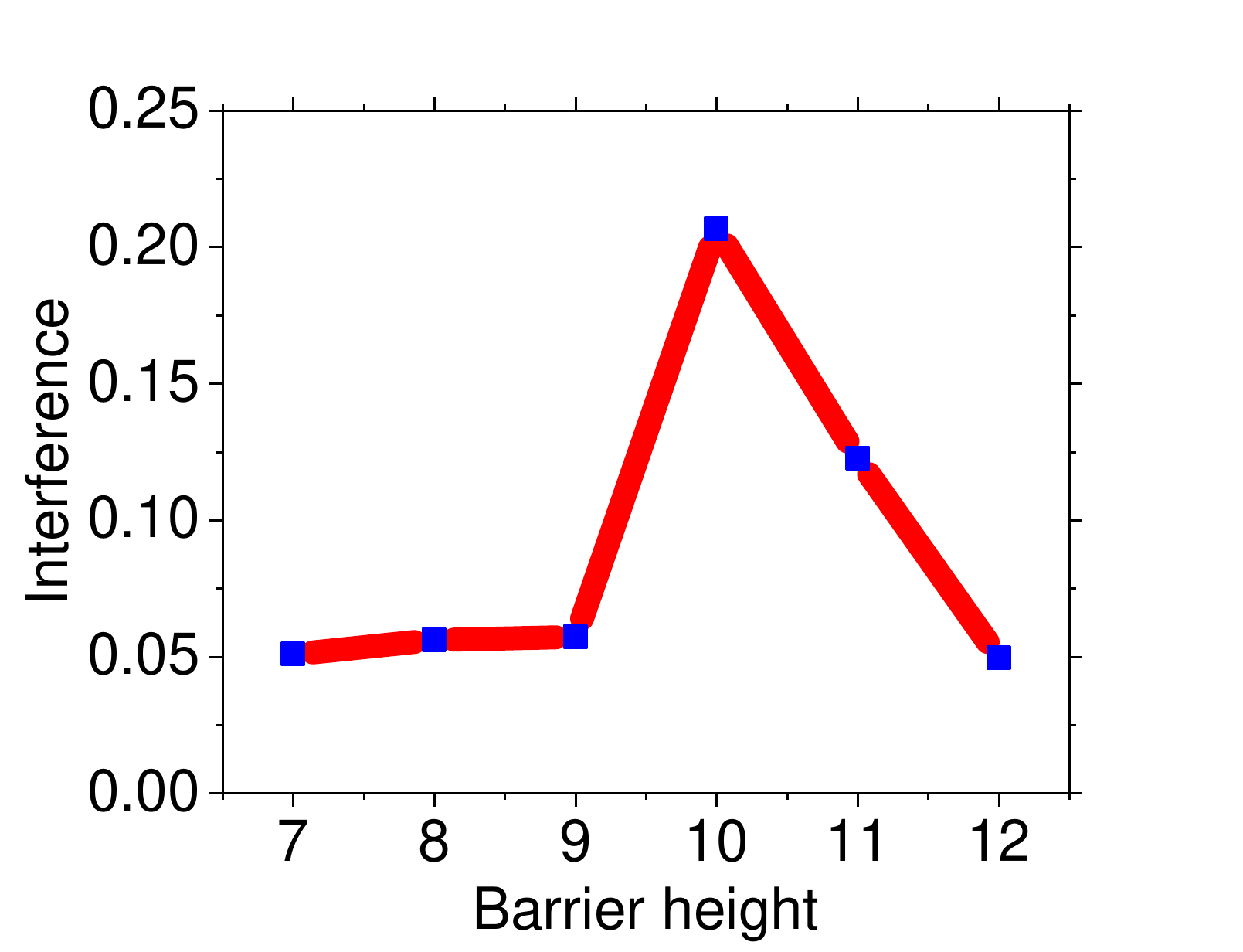}}
\caption{ Interference of  longitudinal and transversal fragmentations. Interference of fragmentations requires three conditions to be satisfied in time, and they are (a) initially fragmented ground state, (b) loss of coherence in the $u$-orbital during initial dynamics, and then (c) build up of coherence in the $u$-orbital.  The  magnitude of interference is defined by the difference between maximal occupancy of the $u$-orbital after building up of coherence and its minimal occupancy.  For the barrier height $V_L<7$, the interference slowly decreases and eventually becomes zero when the $u$-orbital has essentially no initial occupancy.   The  inter-boson interaction is $10\Lambda_0$   and the number of bosons $N=10$.    We show here dimensionless quantities.}
\label{FigS4_new}
\end{figure*}

 Remarkably, at the beginning of the dynamics found for  $V_L\geq 7$, the second natural orbital loses its coherence along with the first natural orbital. Here, the second, third, and fourth natural orbitals are $u$-orbital, excited $g$-orbital, and excited $u$-orbital, respectively.  The loss of coherence of the $u$-orbital occurs due to its  sufficient  initial occupation  at $V_L\geq 7$   and it mimics the trend of the  $g$-orbital. This  loss of coherence of the $u$-orbital is, of course,   a purely many-body phenomenon and it occurs only for the fragmented ground state (also discussed in the main text). If the ground state is initially more fragmented, depending on the barrier height,   the $u$-orbital follows the pattern of loss of coherence for longer times in the process of tunneling.  The $u$-orbital loses its coherence until the moment in time   when there is a swapping of orders of populations between two  higher natural orbitals.   At $V_L=9$ and $10$, we observe that the  $u$-orbital  and   excited $g$-orbital exchange their orders at $t=0.76$ and  2.11, respectively, and afterwards, the $u$-orbital builds up coherence, see the  insets of Figs.~\ref{FigS3} (c) and (d). This build up of coherence in the $u$-orbital defines the interference of the longitudinal and transversal fragmentations, see the main text.
 
 At $V_L=11$, the microscopic mechanism of the fragmentation becomes even richer. Here we find that the build up of coherence in the $u$-orbital is accompanied by the swapping of orders of populations  of excited $g$-orbital  and excited $u$-orbital at $t=4.56$. Further, around $t=17$, the $u$-orbital  and excited $u$-orbital  exchange their orders three times and afterwards, the $u$-orbital builds up coherence. At $V_L=12$, the build up of coherence in the $u$-orbital occurs when the excited $g$-orbital  and  excited $u$-orbital  swap their orders. Also, similar to the fragmentation  mechanism found at $V_L=9$ and 10, the build up of coherence in the $u$-orbital is accompanied by a swapping of the orders of the $u$-orbital and excited $g$-orbital.
 
 At $V_L=16$, as the initial state is essentially fully fragmented and there is   no coupling  between the longitudinal and transversal fragmentations, the $u$-orbital retains the trend  of loss of coherence, also see Fig. 2(d) of the main text.  Accordingly, the populations of the excited $g$-orbital  and excited $u$-orbital  monotonously grow.  All in all, the  interference of the longitudinal and transversal fragmentations occurs when  (i) the $u$-orbital has sufficient initial occupancy, (ii) it  loses its coherency in the initial dynamical evolution,  and then (iii) it  builds up its coherency.  We find that the interference is maximal at $V_L=12$ for the considered inter-boson interaction $\Lambda_0$.

Now,  we discuss  the time-evolution of $\dfrac{n_{{j=2, 3, 4}}(t)}{N}$ for the stronger inter-boson interaction $10\Lambda_0$, see Fig.~\ref{FigS4},  and find how it  leads to a qualitatively richer microscopic mechanism of interference of fragmentations found for the weaker interaction $\Lambda_0$.  We observed   for the weaker interaction that,  whenever the $u$-orbital breaks its natural trend from  loss of coherence to build up of coherence, the $u$-orbital  and excited $g$-orbital  or the excited $g$-orbital  and excited $u$-orbital  interchange their orders.   Here, for $10\Lambda_0$, we find that the orders of the  orbitals interchange multiple times, for example see for $V_L=10$ and $V_L=12$, see Fig.~\ref{FigS4}. Although at the smaller barrier height, say at $V_L=9$, the $u$-orbital  and excited $g$-orbital exchange their orders only once,  but  for $V_L=10$, the $u$-orbital breaks its trend twice from loss of coherence to build up of coherence, at $t=0.66t_{Rabi}$ and $3.60t_{Rabi}$. For the former time, the excited $g$-orbital  and excited $u$-orbital, and for the latter time, the $u$-orbital  and excited $u$-orbital  interchange their orders.  Remarkably, at $V_L=12$, the $u$-orbital  and  excited $g$-orbital interchange their orders twelve times between $t=3t_{Rabi}$ and $6t_{Rabi}$.  Therefore, for strong interaction,  we observe that  the $u$-orbital  breaks its trend of loss of coherence to build up of coherence  multiple times due to the interchange of order by one of three possibilities:  either  the $u$-orbital  and excited $g$-orbital,   or the excited $g$-orbital  and excited $u$-orbital,   or the $u$-orbital  and excited $u$-orbital. Here we find that the transition from loss of coherence to build up of coherence or vice versa is always accompanied by the interchange of orders of higher natural orbitals  until the revival takes place. In the two extreme barrier heights considered, at $V_L=1$,  the $u$-orbital shows transition from build up of coherence to loss of coherence and, at $V_L=16$, from loss of coherence to build up of coherence,  due to the pure effect of the  revival process.

Now, we determine at which barrier height the interference of longitudinal and transversal fragmentations is maximal for the inter-boson interaction $10\Lambda_0$, see Fig.~\ref{FigS4_new}. The interference is calculated from the difference between maximal occupancy of the $u$-orbital  after building up of coherence (but  before the revival process) and its minimal occupancy. We find that at the intermediate barrier heights, from $V_L=7$ to $V_L=12$, the interference of longitudinal and transversal fragmentations take place and, for $10\Lambda_0$, the maximal interference occurs at $V_L=10$. Therefore, with the increase of inter-boson interaction, the maximal interference appears at the lower barrier height, compare to Fig. 2(c) of the main text. 
\clearpage
\section{Dynamics of longitudinal and transversal Position variances}
\subsection{Mean-field dynamics}
\begin{figure*}[!h]
{\includegraphics[trim = 0.1cm 0.5cm 0.1cm 0.2cm, scale=.65]{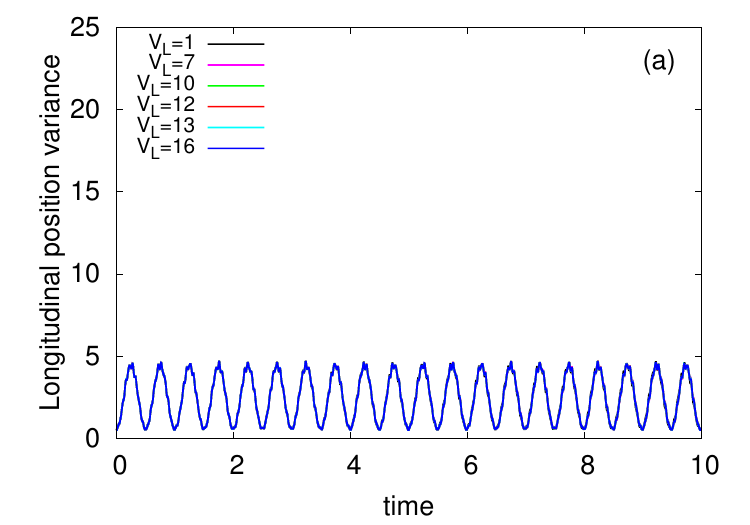}}
{\includegraphics[trim = 0.1cm 0.5cm 0.1cm 0.2cm, scale=.65]{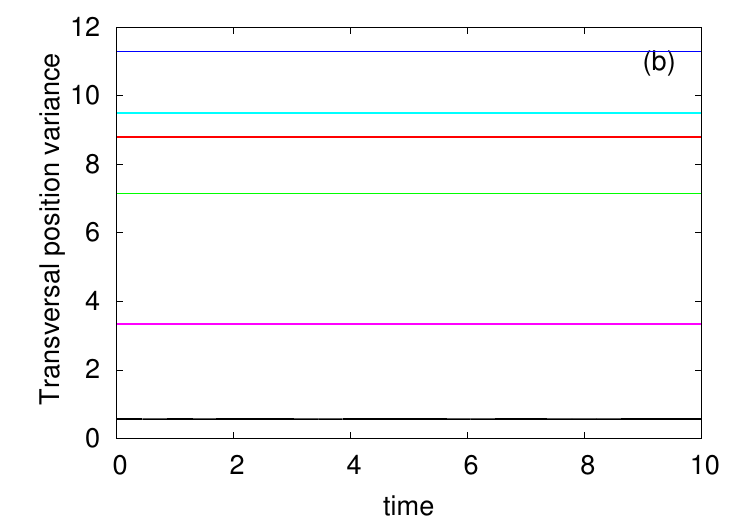}}\\
\caption{Time-dependent many-particle position variances per particle along the longitudinal and transversal directions,  $\dfrac{1}{N}\Delta^2_{\hat{X}}(t)$ and $\dfrac{1}{N}\Delta^2_{\hat{Y}}(t)$, respectively, for different barrier heights. The  inter-boson interaction is $\Lambda_0$. The results are obtained by applying the mean-field theory.  The results for $10\Lambda_0$ essentially overlap those for $\Lambda_0$ and are not plotted. Color codes are explained in the left panel. We show here dimensionless quantities. }
\label{FigS5}
\end{figure*}

Here we start our discussion with the position variance and use it to analyze: (i) how  ground states of different shapes  tunnel when the fragmentation is not considered in the system and (ii) the   Josephson tunneling dynamics in the mean-field limit. Fig.~\ref{FigS5} records the mean-field longitudinal and transversal position variances, $\dfrac{1}{N}\Delta^2_{\hat{X}}(t)$ and $\dfrac{1}{N}\Delta^2_{\hat{Y}}(t)$, respectively, for the inter-particle interaction $\Lambda_0$. Fig.~\ref{FigS5} (a)  shows that the mean-field dynamics of $\dfrac{1}{N}\Delta^2_{\hat{X}}(t)$  is practically independent of the barrier height, $V_L$. Whereas the mean-field $\dfrac{1}{N}\Delta^2_{\hat{X}}(t)$ essentially does not depend on the barrier height for the time durations considered here,  $\dfrac{1}{N}\Delta^2_{\hat{Y}}(t)$   monotonously increases with $V_L$ having almost a frozen dynamical behavior (fluctuation around $10^{-3}$). This monotonous  increase of the base value of  $\dfrac{1}{N}\Delta^2_{\hat{Y}}(t)$  originates   due to the initial deformation in the ground state, see Fig.~\ref{FigS_new}.  For the increased interaction strength  $10\Lambda_0$,  we find that $\dfrac{1}{N}\Delta^2_{\hat{X}}(t)$ and  $\dfrac{1}{N}\Delta^2_{\hat{Y}}(t)$ show practically the same   dynamical behavior as compared to their corresponding results  found for $\Lambda_0$.    Therefore, it suggests that the regime of  the  inter-boson interaction strength considered in this work does not have any qualitative impact on   the mean-field dynamics of the longitudinal and transversal position variances.

\clearpage

\subsection{Many-body dynamics}

\begin{figure*}[!h]
{\includegraphics[trim = 0.1cm 0.5cm 0.1cm 0.2cm, scale=.65]{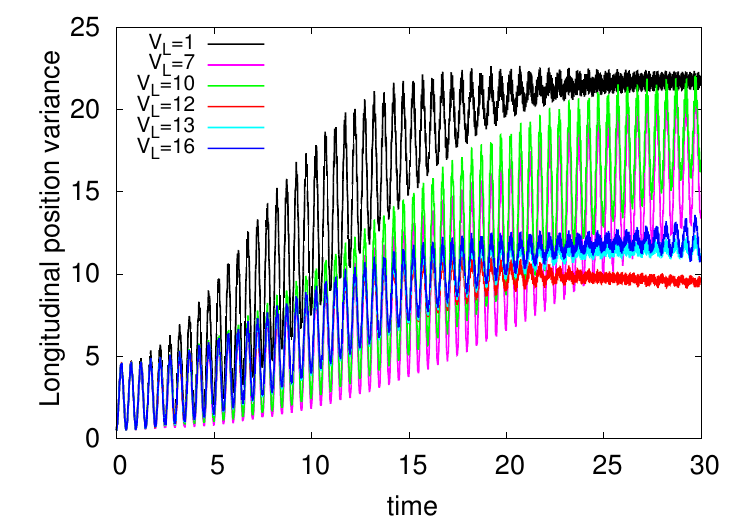}}
\caption{Time-dependent many-body longitudinal position variance per particle,  $\dfrac{1}{N}\Delta^2_{\hat{X}}(t)$,  for different barrier heights. The  inter-boson interaction is $\Lambda_0$   and the number of bosons $N=10$.  See the text for further discussion.  We show here dimensionless quantities.}
\label{FigS6}
\end{figure*}

As the main text demonstrates the  dynamics of the many-body transversal position variance  for the interaction strength $\Lambda_0$, here we discuss the impact of the interference of  longitudinal and transversal fragmentations on the many-body dynamics of the longitudinal position variance, $\dfrac{1}{N}\Delta^2_{\hat{X}}(t)$, see Fig.~\ref{FigS6}. As a general feature, we find that the many-body $\dfrac{1}{N}\Delta^2_{\hat{X}}(t)$ oscillates with a growing amplitude and the rate of growth varies depending on $V_L$. In comparison with the corresponding mean-field dynamics, the maximal deviation of the many-body $\dfrac{1}{N}\Delta^2_{\hat{X}}(t)$ occurs at $V_L=1$ and the minimal at $V_L=12$. We find that the growth of the many-body $\dfrac{1}{N}\Delta^2_{\hat{X}}(t)$ maintains its order from $V_L=1$ to $V_L=10$ which is consistent with the many-body $P(t)$ and the  dynamics of the normalized occupation number, $\eta(t)$, discussed in the main text. For example, if we consider the barrier height until $V_L=10$, the many-body $\dfrac{1}{N}\Delta^2_{\hat{X}}(t)$ shows maximal deviation at $V_L=7$ and minimal at $V_L=10$ in comparison with the respective many-body dynamics at $V_L=1$, which is consistent with $P(t)$ and the normalized occupation number.  By examining Fig.~\ref{FigS6}, we notice that the saturation value of the many-body $\dfrac{1}{N}\Delta^2_{\hat{X}}(t)$ is almost double  for the  fully condensed state (at $V_L=1$) compared to the corresponding saturation value for the fully fragmented state (at $V_L=16$). Moreover, the saturation value of the many-body $\dfrac{1}{N}\Delta^2_{\hat{X}}(t)$ is minimal when the interference of the longitudinal and transversal fragmentations is maximal. Therefore, by analyzing the many-body $\dfrac{1}{N}\Delta^2_{\hat{X}}(t)$ and $\dfrac{1}{N}\Delta^2_{\hat{Y}}(t)$ (in the main text), we observe that the interference of the longitudinal and transversal fragmentations hinders attaining a higher saturation value  for $\dfrac{1}{N}\Delta^2_{\hat{X}}(t)$ and  helps to increase the amplitude of oscillations of  $\dfrac{1}{N}\Delta^2_{\hat{Y}}(t)$.

\begin{figure*}[!h]
{\includegraphics[trim = 0.1cm 0.5cm 0.1cm 0.2cm, scale=.65]{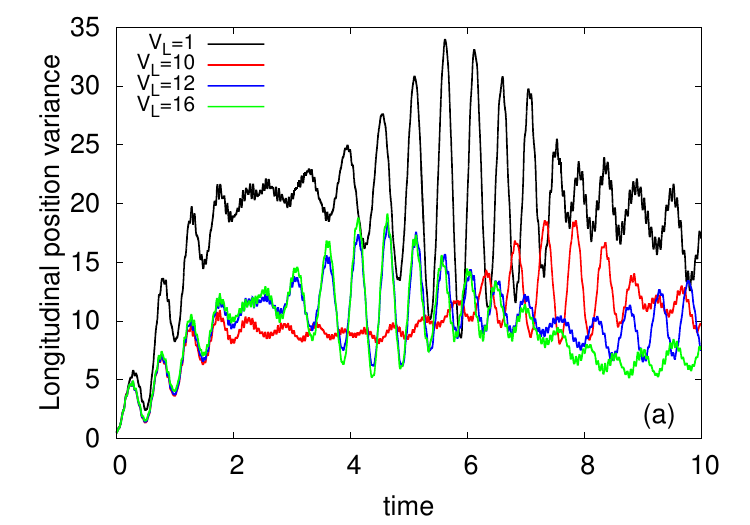}}
{\includegraphics[trim = 0.1cm 0.5cm 0.1cm 0.2cm, scale=.65]{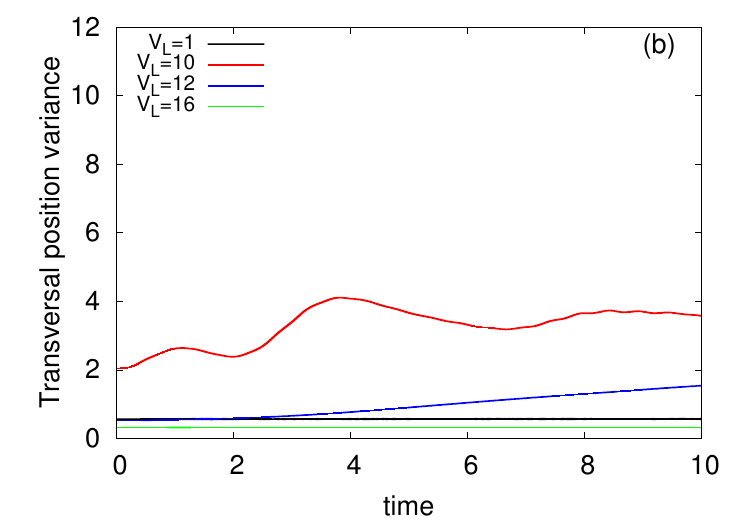}}\\
\caption{Time-dependent many-body position variances per particle along the longitudinal and transversal directions,  $\dfrac{1}{N}\Delta^2_{\hat{X}}(t)$ and $\dfrac{1}{N}\Delta^2_{\hat{Y}}(t)$, respectively, for different barrier heights. The  inter-boson interaction is $10\Lambda_0$   and the number of bosons $N=10$.  See the text for further discussion.  We show here dimensionless quantities. }
\label{FigS7}
\end{figure*}

Now we would demonstrate the many-body longitudinal and transversal position variances for the interaction strength $10\Lambda_0$, see Fig.~\ref{FigS7} (a) and (b), respectively,  and provide a further connection with the respective many-body survival probability, $P(t)$,  and normalized loss of coherence, $\eta(t)$,  discussed in the main text. From  Fig.~\ref{FigS7} (a), we find that the many-body dynamics of $\dfrac{1}{N}\Delta^2_{\hat{X}}(t)$ gradually increases in an oscillatory manner due to the development of longitudinal fragmentation and eventually  it saturates until the revival takes place, see corresponding survival probability plot in Fig.4(a) of main text. The saturation value of  $\dfrac{1}{N}\Delta^2_{\hat{X}}(t)$ is maximal for $V_L=1$, when the initial state is fully condensed, and minimal for $V_L=10$ when the interference of the longitudinal and transversal fragmentations is maximal. The signature of revival shown in the many-body dynamics of $\dfrac{1}{N}\Delta^2_{\hat{X}}(t)$ complements  the time-evolution of the many-body $P(t)$   and normalized loss of coherence, $\eta(t)$, see Fig. 4 of main text. It is found that the revival in the many-body dynamics of $\dfrac{1}{N}\Delta^2_{\hat{X}}(t)$ takes place faster when the system is initially fully fragmented and there is essentially no interference between the longitudinal and transversal fragmentations in the process of tunneling, i.e., at $V_L=16$. Moreover, the process of revival is delayed when the maximal interference of fragmentations occurs.  

Fig.~\ref{FigS7} (b) records the many-body dynamics of $\dfrac{1}{N}\Delta^2_{\hat{Y}}(t)$. For the lowest barrier height $V_L=1$, when the initial ground state is fully condensed, the dynamics of $\dfrac{1}{N}\Delta^2_{\hat{Y}}(t)$ is found to be practically frozen.  As the interference of the longitudinal and transversal fragmentations takes place, we observe oscillatory nature in the dynamics of $\dfrac{1}{N}\Delta^2_{\hat{Y}}(t)$, see for $V_L=10$. Compared to the weak inter-boson  interaction, here we find that the period of oscillations of $\dfrac{1}{N}\Delta^2_{\hat{Y}}(t)$ changes with time due to the combined effect of transversal many-body dynamics  and breathing-motion frequencies. The maximal fluctuations of $\dfrac{1}{N}\Delta^2_{\hat{Y}}(t)$ with a growing amplitude is observed for  $V_L=10$ which goes hand in hand with the maximal coupling of the longitudinal and transversal fragmentations. As we increase the barrier height $V_L>10$, the interference of the longitudinal and transversal fragmentations gradually becomes smaller and the oscillations of $\dfrac{1}{N}\Delta^2_{\hat{Y}}(t)$ slowly decay. Eventually, at $V_L=16$, we observe essentially constant dynamical behavior due to the practically null interference of the longitudinal and transversal fragmentations. Therefore, we come to the conclusion that the detailed investigation of the many-body dynamics of $\dfrac{1}{N}\Delta^2_{\hat{Y}}(t)$, being a sensitive probe of correlations,  encodes the interference of  longitudinal and transversal fragmentations.

\section{Robustness of THE results to the WIDTH OF the INTER-Boson INTERACTION
POTENTIAL}
In the main text, we have made a detailed investigation  on the physics of interference of the  longitudinal and transversal fragmentations in the tunneling process in two spatial dimensions. For our study, we have used for the inter-boson interaction  a Gaussian model potential of finite width, $\sigma=0.25\sqrt{\pi}$. In order  to demonstrate the robustness of our results, we recomputed all the properties discussed in this work for two additional smaller widths, $\sigma=0.25$ and $\sigma=0.25/\sqrt{\pi}$. Note that we purposely do not use the delta-function to  model the inter-boson interaction as the delta-function in two-dimensions does not scatter.

\begin{figure*}[!h]
{\includegraphics[trim = 0.1cm 0.5cm 0.1cm 0.2cm, scale=.25]{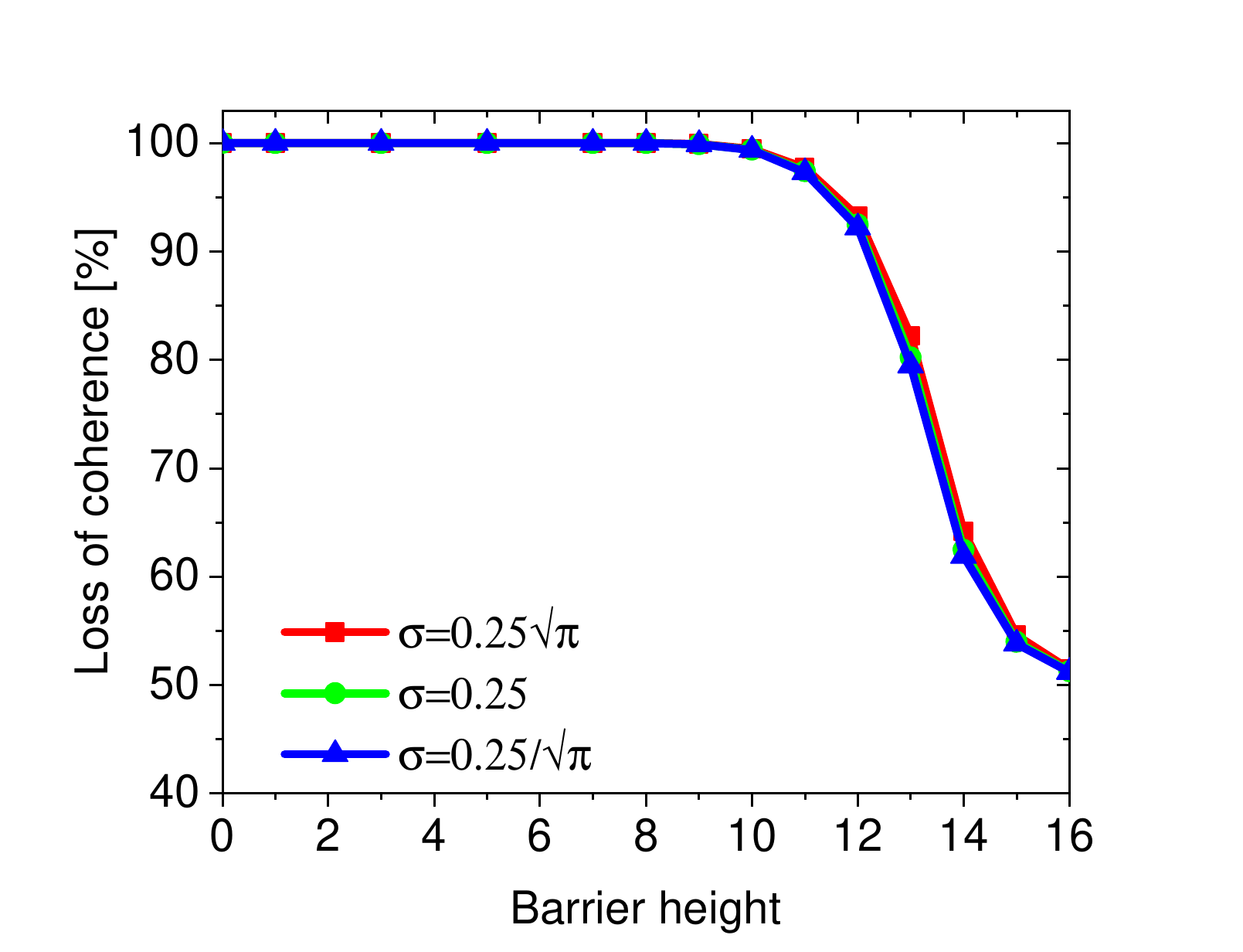}}
{\includegraphics[trim = 0.1cm 0.5cm 0.1cm 0.2cm, scale=.25]{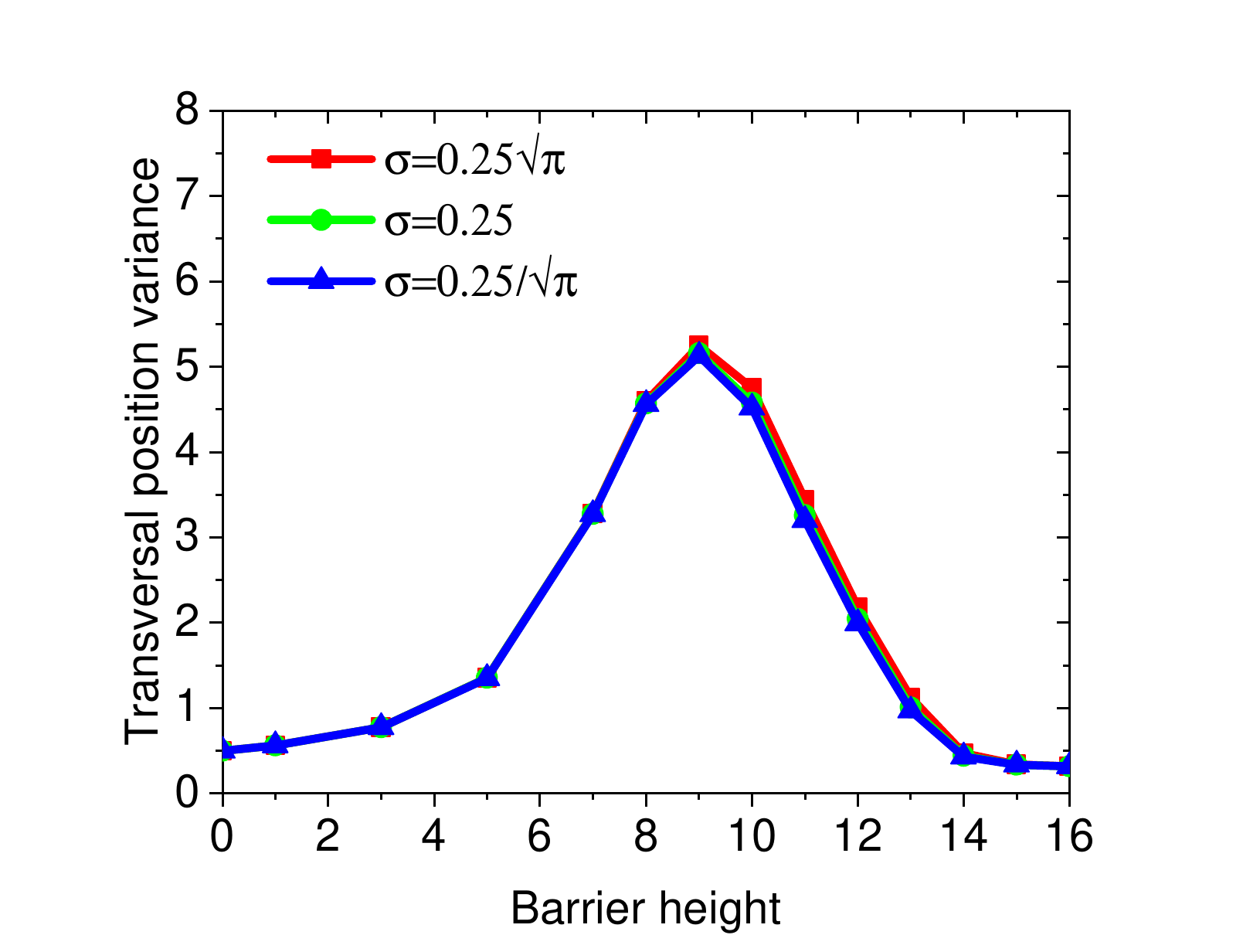}}\\
\caption{(a) Loss of coherence and (b) transversal position variance,
$\dfrac{1}{N}\Delta_{{\hat{Y}}}^2(t)$, for the three different widths of the inter-boson interaction potential, i.e., $\sigma=0.25\sqrt{\pi}$, $0.25$, and $0.25/\sqrt{\pi}$, as a function of the longitudinal barrier height $V_L$. The inter-boson interaction is $\Lambda_0$ and the number of bosons  $N = 10$. We show here dimensionless quantities.}
\label{FigS8}
\end{figure*}

Here, we begin with the discussion of static properties, i.e, the loss of coherence and transversal position variance as a function of the longitudinal barrier height for the widths of the Gaussian model potential $\sigma=0.25\sqrt{\pi}$, $0.25$, and $0.25/\sqrt{\pi}$, see Fig.~\ref{FigS8}. We find that the qualitative physics of the static properties, presented in this work, does not depend on $\sigma$. Moreover, the difference between the results, found for $\sigma=0.25$ and $0.25/\sqrt{\pi}$, is smaller compared to the respective difference obtained for $\sigma=0.25\sqrt{\pi}$ and  $0.25$. This suggests  that  decreasing of the width, $\sigma$, slowly reduces the quantitative difference of a particular quantity, at least for the properties examined here.

\begin{figure*}[!h]
{\includegraphics[trim = 0.1cm 0.5cm 0.1cm 0.2cm, scale=.65]{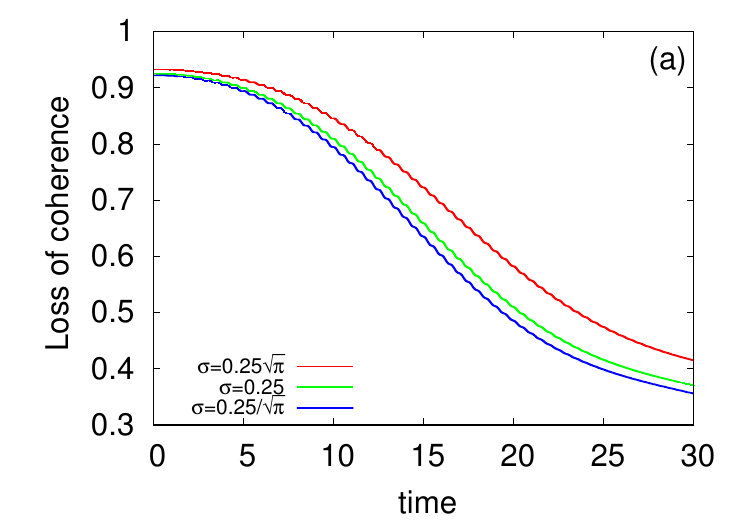}}
{\includegraphics[trim = 0.1cm 0.5cm 0.1cm 0.2cm, scale=.65]{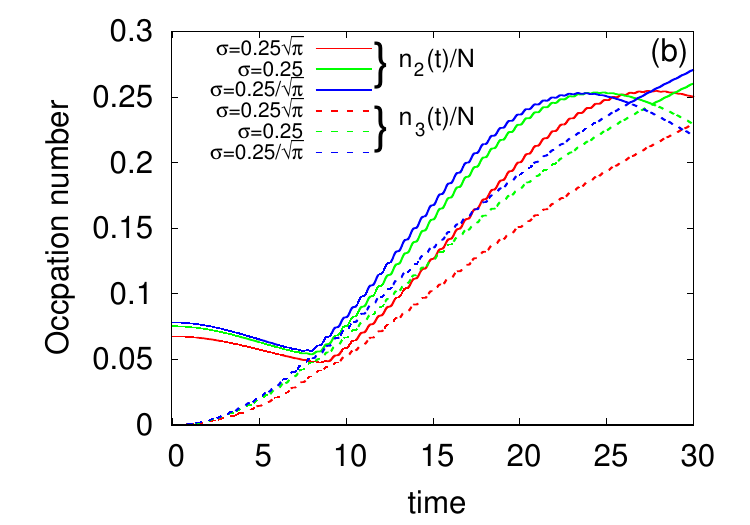}}\\
\vspace{0.5cm}
{\includegraphics[trim = 0.1cm 0.5cm 0.1cm 0.2cm, scale=.65]{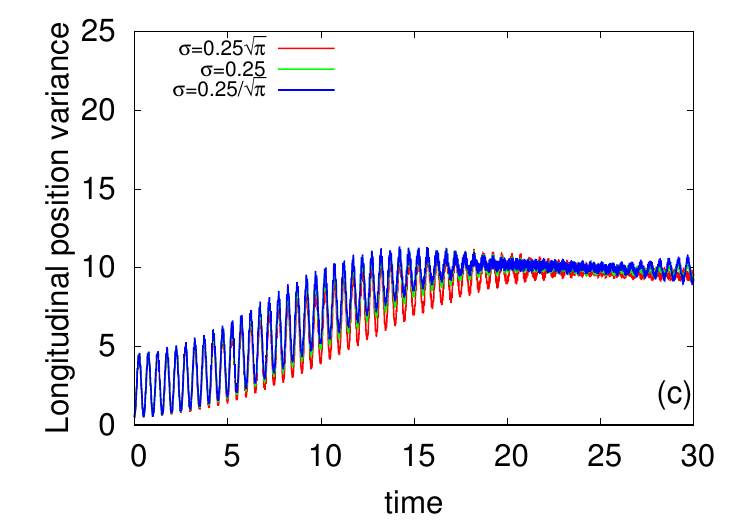}}
{\includegraphics[trim = 0.1cm 0.5cm 0.1cm 0.2cm, scale=.65]{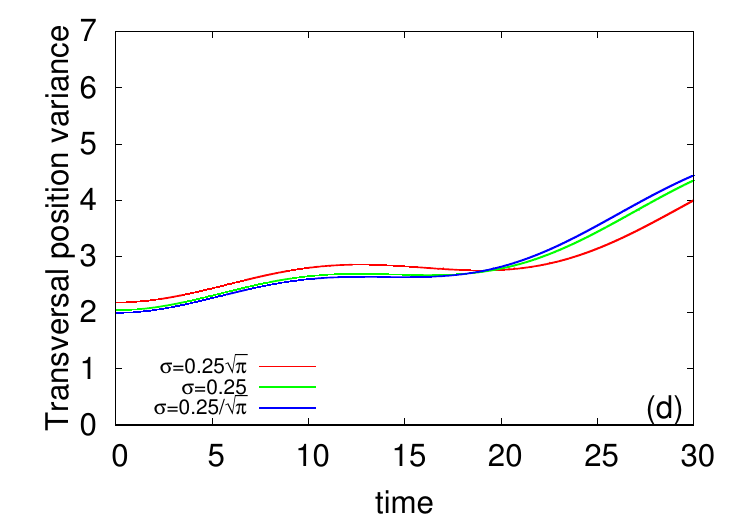}}\\
\caption{ Dependence of the many-body dynamics of the initially-prepared  ground state at the barrier height $V_L=12$  on  the three different widths of the inter-boson interaction potential, i.e., $\sigma=0.25\sqrt{\pi}$, $\sigma=0.25$, and $\sigma=0.25/\sqrt{\pi}$. The dynamics is represented by  the (a) loss of coherence, (b) occupation of the second and third natural orbitals, (c) longitudinal position variance, and (d) transversal position variance. The  inter-boson interaction is  $\Lambda_0$ and the number of bosons $N=10$.  In panel (b), the solid and dashed lines represent the occupation of the second, $n_2(t)/N$,  and third, $n_3(t)/N$,  natural orbitals, respectively.  We show here dimensionless quantities.}
\label{FigS9}
\end{figure*}

  To present the robustness of the width,  $\sigma$, at the many-body dynamics, we select the dynamics of the ground state at $V_L=12$ when the interference of the longitudinal and transversal fragmentations is maximal.  Fig.~\ref{FigS9} depicts the many-body dynamics of  loss of coherence, $\dfrac{n_1(t)}{N}$, occupations of the second and third natural orbitals, $\dfrac{n_2(t)}{N}$ and $\dfrac{n_3(t)}{N}$, respectively, longitudinal position variance, $\dfrac{1}{N}\Delta_{{\hat{X}}}^2(t)$,  and  transversal position variance, $\dfrac{1}{N}\Delta_{{\hat{Y}}}^2(t)$, at $V_L=12$ for the widths  $\sigma=0.25\sqrt{\pi}$, $0.25$, and $0.25/\sqrt{\pi}$.  First,  it is found that $\dfrac{n_1(t)}{N}$ is the $g$-orbital throughout the tunneling process for all $\sigma$. Also, we observe that the $g$-orbital loses coherence comparatively quicker (slower)  for $\sigma=0.25/\sqrt{\pi}$ ($\sigma=0.25\sqrt{\pi})$ which implies faster (slower) development of fragmentation as shown in Fig.~\ref{FigS9} (a). The unchanged qualitative feature of the occupancy of the $g$-orbital  suggests its robustness with the width of the inter-boson interaction.

   Fig.~\ref{FigS9} (b) presents the occupation  of the  second and third natural orbitals. For $\sigma=0.25\sqrt{\pi}$, discussed in the main text, the second and third natural orbitals remain as $u$-orbital   and excited $g$-orbital, and the build up of coherence in the $u$-orbital manifests  the  interference of  longitudinal and transversal fragmentations. Here, also for the widths $\sigma=0.25$ and $0.25/\sqrt{\pi}$, we observe a build up of coherence in the $u$-orbital with the faster rate for the width $\sigma=0.25/\sqrt{\pi}$. It is found that the qualitative features of occupations of the second and third natural orbitals are the  same for all $\sigma$ until a swapping between the orders of $u$-orbital  and excited $g$-orbital happens for $\sigma=0.25$ and $0.25/\sqrt{\pi}$. This swapping of orders of the orbitals  would also happen for  $\sigma=0.25\sqrt{\pi}$ if one would compute the dynamics for a longer time than 30 Rabi Cycles. Therefore, Fig.~\ref{FigS9} (b) exhibits the robustness of the interference of longitudinal and transversal fragmentations to the width of the inter-boson interaction potential.

   As the variances are  sensitive probe of correlations, we find here that the developed fragmentation  in the system only quantitatively influences the many-body $\dfrac{1}{N}\Delta_{{\hat{X}}}^2(t)$  and $\dfrac{1}{N}\Delta_{{\hat{Y}}}^2(t)$ for different $\sigma$. Also, Figs.~\ref{FigS9} (c) and (d)  attribute that decreasing the width $\sigma$ of the inter-boson interaction potential gradually diminishes the quantitative difference between the dynamics  of a given quantity for the studies made here. All in all, we  observe that the width of the inter-boson interaction potential does not qualitatively impact the many-body  physics of the interference of fragmentations we focused on and discussed in this work.
  
Further, we have computed all the quantities  for each of the  barrier heights for the inter-boson interaction strength $10\Lambda_0$ with the two additional smaller widths,  $\sigma=0.25$ and $0.25/\sqrt{\pi}$. We have checked and verified that the qualitative physics of the dynamics of the different  fragmented states is independent of $\sigma$ for the stronger interaction $10\Lambda_0$ as well, and thus they are not shown graphically. 

Finally,  in the mean-field dynamics, we notice that the time-evolution of  all the quantities obtained for $\sigma=0.25$ and $\sigma=0.25/\sqrt{\pi}$  fall on top of the respective results computed for $\sigma=0.25\sqrt{\pi}$ (not shown). Therefore, the width of the inter-boson interaction does  essentially not influence the mean-field dynamics presented in this work, which is interesting for itself.

\section{Robustness of the results to the form of the boson-boson interaction: The case of dipolar interaction}
\begin{figure*}[!h]
{\includegraphics[trim = 0.1cm 0.5cm 0.1cm 0.2cm, scale=.60]{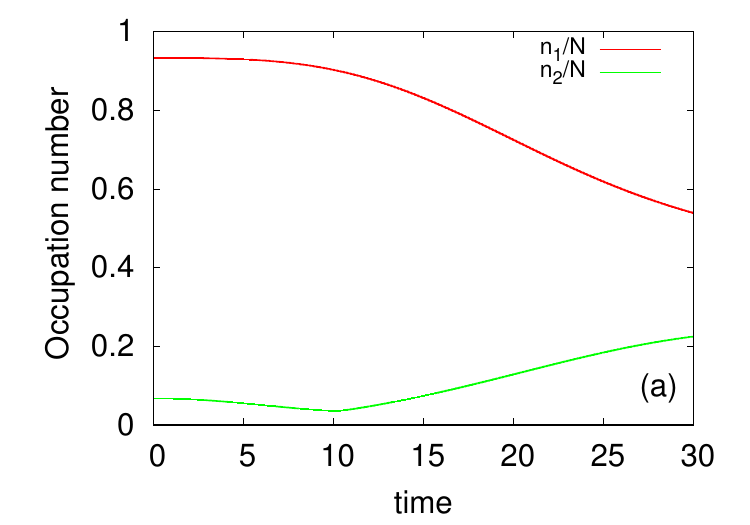}}
{\includegraphics[trim = 0.1cm 0.5cm 0.1cm 0.2cm, scale=.60]{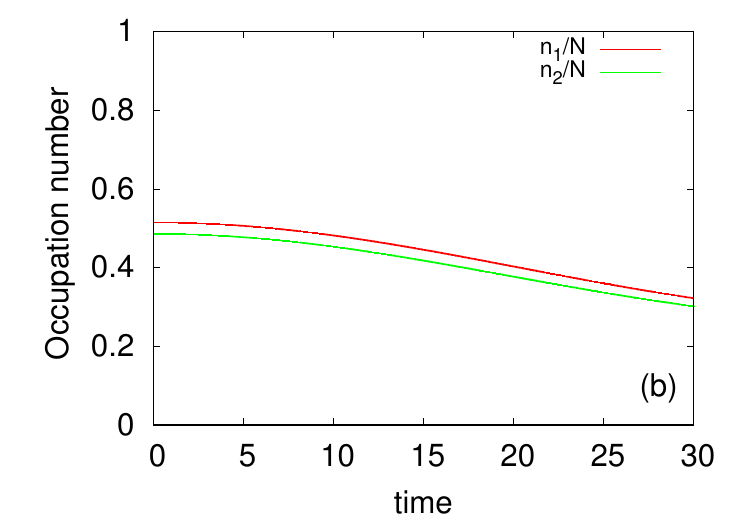}}\\
\caption{ Dynamics for dipolar interaction. Occupations of the first and second natural orbitals, $\dfrac{n_1(t)}{N}$ and $\dfrac{n_2(t)}{N}$, respectively, for (a) $V_L=12$ and (b) $V_L=16$.  The  inter-boson interaction is  $\Lambda_0$ and the number of bosons $N=10$.   Robustness of the interference of fragmentations to the form of the inter-boson interaction is found.  We show here dimensionless quantities.}
\label{FigS_new_1}
\end{figure*}

This section demonstrates the robustness of the interference of longitudinal and transversal fragmentations to the form of the boson-boson interaction. In the main text and throughout the supplemental material, the boson-boson interaction is chosen as a  Gaussian form. Here, we investigate the interference of fragmentations for  a different shape of the interaction, say, the dipolar interaction which can be written mathematically as, $W(\textbf{r}_j-\textbf{r}_k)=\dfrac{\lambda_0}{|r_j-r_k|^3+\Delta^3}$, where $\Delta^3=0.07$ is the threshold of the long-range interaction \cite{Dutta2019_s}. Such long-range interaction is relevant for atomic clouds made of Cr,  Dy,  or  Er.  Here also, $\lambda_0$ is the interaction strength which is related to the interaction parameter $\Lambda_0=\lambda_0(N-1)$.  Now, we consider the same number of bosons and interaction strength as described in Figure 2(c) and 2(d) of the manuscript. Fig.~\ref{FigS_new_1} presents the occupation of the first and second  natural orbitals for the barrier heights  $V_L=12$ and $V_L=16$.  We observe that the  dynamics of $\dfrac{n_1(t)}{N}$ and $\dfrac{n_2(t)}{N}$  are  qualitatively  the same as described in the main text. For intermediate hight of  $V_L=12$, here also  we find loss of coherence followed by build up of coherence in the second natural orbital. This investigation exhibits the robustness of the interference of fragmentations to the form of the inter-boson interaction.

\section{Details of CONVERGENCES OF QUANTITIES in the tunneling dynamics}
The multiconfigurational time-dependent Hartree for bosons (MCTDHB) method  is used in the present work to compute the  ground (initial)  state which eventually becomes fragmented depending on the barrier height $V_L$.  To compute the ground state and its subsequent real-time propagation,  the many-body Hamiltonian is represented by $128^2$ exponential discrete-variable-representation
grid points  in a box of size $[-10, 10)\times [-10, 10)$. In the main text, we calculated the many-body quantities with  $M=8$ time-adaptive orbitals. Here we provide the convergences of the quantities discussed in this work, i.e.,  we show that the time-dependent many-boson wavefunction built from $M=8$ time-adaptive orbitals leads to numerically converged results.   We have verified the convergences of the quantities with $M=10$ time-adaptive orbitals using $128^2$ exponential discrete-variable-representation grid points and also  with $M=8$  time-adaptive orbitals using increased grid density of  $256^2$ exponential discrete-variable-representation grid points for all the barrier heights and  the two interaction strengths $\Lambda_0$ and $10\Lambda_0$. In order to demonstrate the convergences, we select the barrier height $V_L=13$, the maximal barrier height for which the interference of fragmentations is   appreciable, for the interaction strength $\Lambda_0$.

\begin{figure*}[!h]
{\includegraphics[trim = 0.1cm 0.5cm 0.1cm 0.2cm, scale=.60]{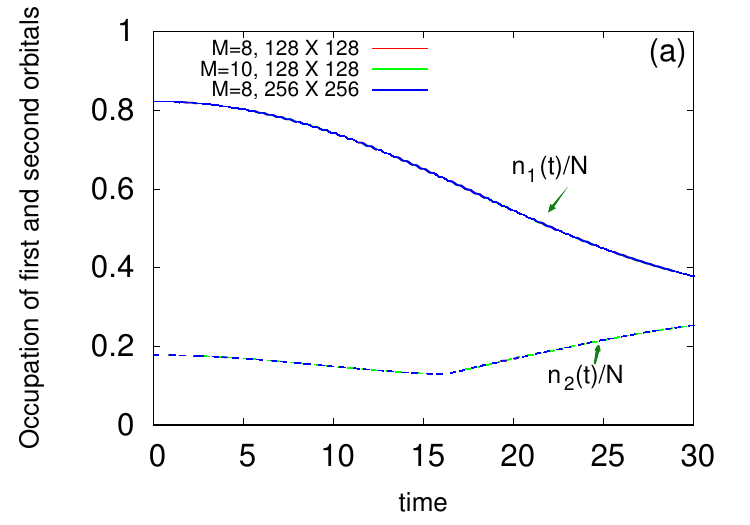}}
{\includegraphics[trim = 0.1cm 0.5cm 0.1cm 0.2cm, scale=.60]{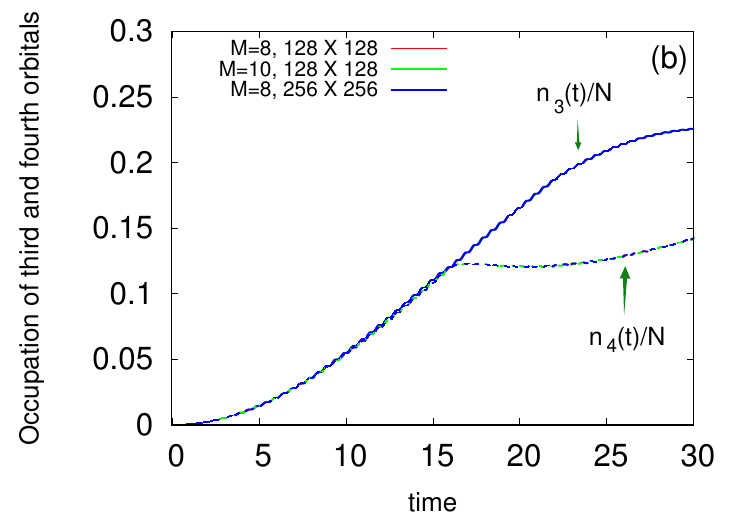}}\\
\vspace{0.5cm}
{\includegraphics[trim = 0.1cm 0.5cm 0.1cm 0.2cm, scale=.60]{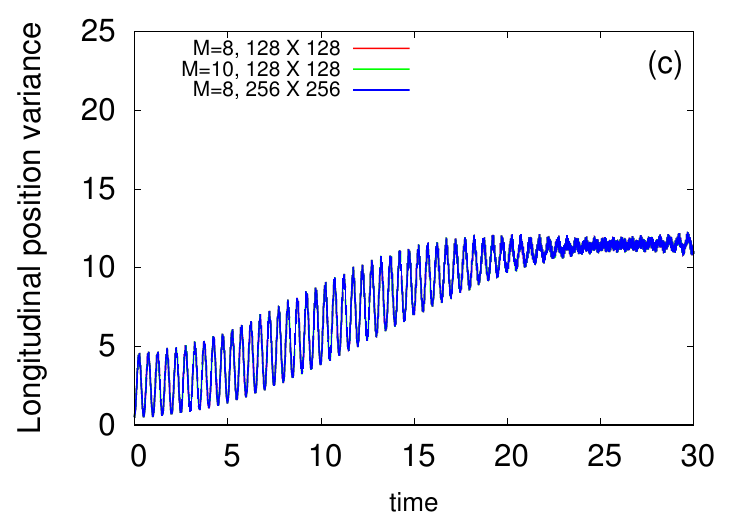}}
{\includegraphics[trim = 0.1cm 0.5cm 0.1cm 0.2cm, scale=.60]{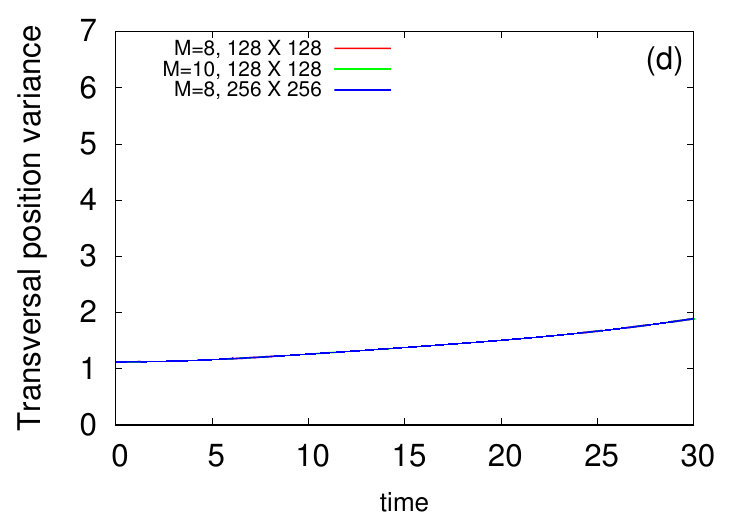}}\\
\caption{ Convergences of the  (a)  occupations of the first and second natural orbitals, (b)  occupations of the third and fourth natural orbitals, (c) many-body longitudinal position variance, and (d) many-body transversal position variance at the barrier height  $V_L=13$   with the number of orbitals and grid points. In panel (a)[(b)], the solid lines represent  the first [third] natural orbital and the dashed lines depict the second [fourth] natural orbital. The  inter-boson interaction is  $\Lambda_0$ and the number of bosons $N=10$. Convergences are verified using $M=10$ time-adaptive orbitals with $128 \times 128$ grid points and  $M=8$ time-adaptive orbitals with $256 \times 256$ grid points. We show here dimensionless quantities.}
\label{FigS10}
\end{figure*}

Fig.~\ref{FigS10} represents the occupations of the first, second, third, and fourth natural orbitals,  longitudinal position variance, $\dfrac{1}{N}\Delta_{{\hat{X}}}^2(t)$,  and  transversal position variance, $\dfrac{1}{N}\Delta_{{\hat{Y}}}^2(t)$, for  $V_L=13$. The overlapping curves for all quantities  with increasing grid density and number of orbitals signify that the results presented in this work are very well converged for $M=8$ time-adaptive orbitals with $128^2$ exponential discrete-variable-representation grid points in a box of size $[-10, 10)\times [-10, 10)$. 

\clearpage



\begin{thebibliography}{}
\bibitem{Dalfovo1999}
F.  Dalfovo,  S.  Giorgini, L. P.  Pitaevskii,  and S. Stringari,    Rev. Mod. Phys. \textbf{71},  463 (1999). 
 \bibitem{Cederbaum2007}
L. S. Cederbaum, A.  I. Streltsov, Y.  B. Band, and O.  E. Alon,
Phys. Rev. Lett. \textbf{98}, 110405 (2007).

\bibitem{Alon2005}
O.  E. Alon, A.  I. Streltsov, and L.  S. Cederbaum, 
Phys. Rev. Lett. \textbf{95}, 030405 (2005).
\bibitem{Bhowmik2016}
A.  Bhowmik, P.  K.  Mondal, S.  Majumder, and B.  Deb, 
Phys. Rev. A \textbf{93}, 063852 (2016).
\bibitem{Schmidt2022}
M.  Schmidt, L.  Lassabli\`ere, G.  Qu\'em\'ener, and T.  Langen, 
Phys. Rev. Research \textbf{4}, 013235   March (2022).

 \bibitem{Bloch2008}
I. Bloch, J. Dalibard, and W. Zwerger, Rev. Mod. Phys.
\textbf{80}, 885 (2008).


 \bibitem{Burchinati2017}
A. Burchinati,  C.  Fort,  and M.  Modugno,        Phys. Rev. A \textbf{95},  023627 (2017).
 \bibitem{Levy2007}
S. Levy, E. Lahoud, I.  Shomroni,  and J.  Steinhauer,  Nature (London) \textbf{449}, 579 (2007).
\bibitem{Smerzi1997}
A.   Smerzi,   S.  Fantoni,  S.  Giovanazzi,   and S. R. Shenoy,     Phys. Rev. Lett.  \textbf{79},  4950  (1997).
\bibitem{Albiez2005}
M. Albiez, R. Gati, J. F\"olling, S. Hunsmann, M. Cristiani, M. K. Oberthaler, Phys. Rev. Lett. \textbf{95}, 010402
(2005).
\bibitem{Milburn1997}
G. J. Milburn, J. Corney, E. M. Wright, D. F. Walls,
Phys. Rev. A \textbf{55}, 4318 (1997).
\bibitem{Schumm2005}
 T. Schumm, S. Hofferberth, L. M. Andersson, S. Wildermuth, S. Groth, I. Bar-Joseph, J. Schmiedmayer, and P.
Kr\"uger, Nature Physics \textbf{1}, 57 (2005).

\bibitem{Orzel2001}
C. Orzel, A. K. Tuchman, M. L. Fenselau, M. Yasuda,
and M. A. Kasevich, Science \textbf{291}, 2386 (2001).
\bibitem{Wu2022}
Q. Wu, L. Mancino, M. Carlesso, M. A. Ciampini, L. Magrini, N. Kiesel, and M. Paternostro, PRX Quantum \textbf{3}, 010322 (2022).
\bibitem{Zibold2010}
T. Zibold, E. Nicklas, C. Gross, and M.  K. Oberthaler, 
Phys. Rev. Lett. \textbf{105}, 204101 (2010).
\bibitem{Abbarchi2013}
M. Abbarchi, A.  Amo, V. G.  Sala, D. D. Solnyshkov, H. Flayac,
L. Ferrier, I. Sagnes, E.  Galopin, A. Lemaitre, G.  Malpuech, and J. Bloch, Nature Physics \textbf{9}, 275  (2013).
\bibitem{Valtolina2015}
G. Valtolina, A. Burchianti, A. Amico, E.  Neri, K.  Xhani, J.  A.  Seman, A. Trombettoni, A.  Smerzi, M.  Zaccanti, M.
 Inguscio, and G.  Roati, Science  \textbf{350},  1505 (2015).
\bibitem{Hou2018}
J.  Hou, X. -W. Luo, K.  Sun, T.  Bersano, V.  Gokhroo, S.  Mossman, P.  Engels, and C.  Zhang,  Phys. Rev. Lett.  \textbf{120},  120401  (2018).

\bibitem{Nozieres1982}
P. Nozi\'eres and  D. Saint James,   J. Phys.  \textbf{43}, 1133 (1982).

\bibitem{Spekkens1999}
R. W. Spekkens and J. E.  Sipe,   Phys. Rev. A  \textbf{59}, 3868 (1999).

\bibitem{Mueller2006}
E. J. Mueller, T. -L. Ho, M. Ueda, and G. Baym
Phys. Rev. A \textbf{74}, 033612 (2006).
\bibitem{Bader2009}
P. Bader and U. R.  Fischer,   Phys. Rev. Lett.  \textbf{103}, 060402 (2009).
\bibitem{Zhou2013}
Q. Zhou  and X. Cui,  Phys. Rev. Lett.  \textbf{110}, 140407 (2013).
\bibitem{Kang2014}
 M. -K. Kang  and U. R.  Fischer, Phys. Rev. Lett.  113, 140404 (2014).
\bibitem{Lode2017}
A. U. J. Lode  and C.  Bruder,      Phys. Rev. Lett. \textbf{118},  013603   (2017).
\bibitem{Chatterjee2020}
B. Chatterjee, C.  L\'ev\^eque,  J. Schmiedmayer,  and A. U. J. Lode, Phys. Rev. Lett. \textbf{125}, 093602 (2020).
\bibitem{Sakmann2009}
K.  Sakmann,  A. I.   Streltsov,  O. E.  Alon,  and L. S.  Cederbaum,   Phys. Rev. Lett. \textbf{103},   220601 (2009).
 \bibitem{Vargas2021}
J. Vargas, M.  Nuske,  R.  Eichberger,  C.  Hippler,  L.  Mathey,  and  A.   Hemmerich,    Phys. Rev. Lett. \textbf{126},  200402 (2021).
 
 \bibitem{Erdmann2018}
J. Erdmann, S. I.  Mistakidis,  and P.  Schmelcher,  Phys. Rev. A \textbf{98}, 053614 (2018).
\bibitem{Theel2020}
F. Theel, K.  Keiler, S. I.  Mistakidis,  and  P. Schmelcher,  New J. Phys. \textbf{22}, 023027 (2020).
 \bibitem{Sias2007}
C. Sias,  A.  Zenesini,  H. Lignier,  S.  Wimberger,  D.  Ciampini,  O.  Morsch,  and  E.  Arimondo,    Phys. Rev. Lett. \textbf{98}, 120403 (2007).

\bibitem{Fialko2012}
O. Fialko, A. S. Bradley, and J. Brand,  Phys. Rev. Lett. \textbf{108}, 015301
(2012).
\bibitem{Spagnolli2017}
G. Spagnolli, G. Semeghini, L. Masi, G. Ferioli, A. Trenkwalder, S. Coop, M. Landini, L. Pezz\`e, G. Modugno, M. Inguscio, A. Smerzi, and M. Fattori, Phys. Rev. Lett. \textbf{118}, 230403  (2017).
\bibitem{Burchianti2018}
A. Burchianti, F. Scazza, A. Amico, G. Valtolina, J. A. Seman, C. Fort, M. Zaccanti, M. Inguscio, and  G. Roati, Phys. Rev. Lett. \textbf{120}, 025302 (2018).
\bibitem{Xhani2020}
K. Xhani, E. Neri, L. Galantucci, F. Scazza, A. Burchianti, K. -L. Lee, C. F. Barenghi, A. Trombettoni, M. Inguscio, M. Zaccanti, G. Roati, and N. P. Proukakis, Phys. Rev. Lett. \textbf{124}, 045301 (2020).
\bibitem{Bhowmik2020}
A.  Bhowmik, S.  K.  Haldar,  and O.  E. Alon,  Sci.  Rep.  \textbf{10},   21476  (2020).


\bibitem{Bhowmik2022}
A.  Bhowmik  and  O. E.  Alon,     Sci.  Rep.  \textbf{12},   627  (2022).

\bibitem{Christensson2009}
J. Christensson, C.  Forss\'en, S. \AA{}berg, S. M.  Reimann, 
Phys. Rev. A \textbf{79}, 012707 (2009).
\bibitem{Doganov2013}
R. A.  Doganov,  S.   Klaiman,  O. E.   Alon,  A. I.   Streltsov, and  L. S.  Cederbaum,    Phys. Rev. A \textbf{87},  033631  (2013).

\bibitem{Fischer2015}
U. R. Fischer, A. U. J. Lode, and B. Chatterjee,  Phys. Rev. A \textbf{91}, 063621 (2015).
\bibitem{Supplement}
Supplemental materials of this work.
\bibitem{Streltsov2007}
A. I. Streltsov, O. E. Alon, and L. S. Cederbaum
Phys. Rev. Lett. \textbf{99}, 030402 (2007).
\bibitem{Alon2008}
O. E. Alon, A. I.  Streltsov,  and L. S.  Cederbaum,    Phys. Rev. A \textbf{77},   033613  (2008).

\bibitem{Nguyen2019}
J. H. V. Nguyen,  M. C.  Tsatsos,  D. Luo, A. U. J.  Lode, G. D. Telles,  V. S.  Bagnato,  and R. G. Hulet,  Phys. Rev. X \textbf{9}, 011052 (2019).
\bibitem{Lode2020}
A. U. J. Lode, C. L\'ev\^eque, L. B. Madsen, A. I. Streltsov, and O.  E.  Alon,  Rev. Mod. Phys. \textbf{92}, 011001 (2020).

\bibitem{Klaiman2015}
S. Klaiman  and O.  E. Alon,  Phys. Rev. A \textbf{91}, 063613 (2015).
\bibitem{Kronke2013}
S. Kr\"onke, L. Cao, O. Vendrell, and P.  Schmelcher,  New J. Phys.  \textbf{15}, 063018 (2013).

\bibitem{Cao2013}
L. Cao, S. Kr\"onke, O. Vendrell, and  P. Schmelcher,  J. Chem. Phys. \textbf{139}, 134103  (2013). 

 \bibitem{Chen2018}
J. Chen, J. M. Schurer, and P. Schmelcher,  Phys. Rev. Lett. \textbf{121}, 043401 (2018).
  \bibitem{Schurer2017}
J. M. Schurer, A. Negretti, and P. Schmelcher, Phys. Rev. Lett. \textbf{119}, 063001 (2017).

  \bibitem{Mistakidis2019}
S. I.  Mistakidis, G. C. Katsimiga, G. M. Koutentakis, Th.  Busch,  and P. Schmelcher,  Phys. Rev. Lett. \textbf{122}, 183001  (2019).


 
\bibitem{Capello2007}
M. Capello, F. Becca, M. Fabrizio, and S. Sorella,  Phys. Rev. Lett. \textbf{99}, 056402 (2007).
 \bibitem{Dunjko2001}
V. Dunjko, V. Lorent, and M. Olshanii,  Phys. Rev. Lett. \textbf{86}, 5413 (2001).

\bibitem{Paredes2004}
B. Paredes, A. Widera, V. Murg, O. Mandel, S. F\"olling, I. Cirac, G. V. Shlyapnikov, T. W. H\"ansch, and I. Bloch,   Nature (London) \textbf{429} 277 (2004).


\bibitem{Amico2022}
L.  Amico, D.  Anderson, M.  Boshier, J. -P.  Brantut, L. -C. Kwek, A.  Minguzzi, and W. von Klitzing, Rev. Mod. Phys. \textbf{94}, 041001 (2022).
\bibitem{Baak2024}
J.-G.  Baak and U. R. Fischer, Phys. Rev. Lett. \textbf{132}, 240803 (2024).


  
 
\end{thebibliography}

\begin{thebibliography}{}
\bibitem{Alon2019b_s}
  O.  E. Alon,   Analysis of a Trapped Bose-Einstein Condensate in Terms of Position, Momentum, and Angular-Momentum Variance.   Symmetry \textbf{11},   1344 (2019).
\bibitem{Lode2020_s}
A.  U.  J. Lode,   C.  L\'ev\^eque,  L. B.  Madsen,   A. I.  Streltsov, and  O. E.    Alon,     Colloquium: Multiconfigurational time-dependent Hartree approaches for indistinguishable particles.  Rev.  Mod. Phys. \textbf{92},  011001 (2020).
\bibitem{Dutta2019_s}
S.  Dutta, M.  C.  Tsatsos, S.  Basu, and  A.  U. J.  Lode, Management of the correlations of UltracoldBosons in triple wells. 
 New Journal of Physics \textbf{21}, 053044 (2019).


\end{thebibliography}

\end{document}